\documentclass[%
 reprint,
unsortedaddress,
 amsmath,amssymb,
 aps,
 pra,
floatfix,
]{revtex4-1}

\usepackage{graphicx}
\usepackage{dcolumn}
\usepackage{bm}
\usepackage{wrapfig}
\usepackage{subfig}
\usepackage{epsfig}
\usepackage{epstopdf}
\usepackage{tikz}
\usepackage{tkz-graph}
\usepackage{natbib}

\begin{document}

\preprint{APS/123-QED}

\title{Vulnerability of Networks Against Critical Link Failures}

\author{Serdar \c{C}olak}
 \affiliation{Bo\~gazi\c{c}i University, Dept. of Civil Engineering}
 \email{Corresponding Author, serdar.colak@boun.edu.tr}

\author{Hilmi Lu\c{s}}
    \affiliation{Bo\~gazi\c{c}i University, Dept. of Civil Engineering}

\author{Ali Rana At\i lgan}
 \affiliation{Sabanc\i \space  University, School of Engineering and Natural Sciences}

\begin{abstract}
Networks are known to be prone to link failures. In this paper we set out to investigate how networks of varying connectivity patterns respond to different link failure schemes in terms of connectivity, clustering coefficient and shortest path lengths. We then propose a measure, which we call the vulnerability of a network, for evaluating the extent of the damage these failures can cause. Accepting the disconnections of node pairs as a damage indicator, vulnerability simply represents how quickly the failure of the critical links cause the network to undergo a specified damage extent. Analyzing the vulnerabilities under varying damage specifications shows that scale free networks are relatively more vulnerable for small failures, but more efficient; whereas Erd\"{o}s-R\'{e}nyi networks are the least vulnerable despite lacking any clustered structure.
\end{abstract}

\maketitle

\section{\label{sec:level1}Introduction}

Although the definition of network failure varies, the undisputed
fact is that nodes or links can and will fail. The question of how
vulnerable networks are against random failures or targeted attacks
has been investigated by numerous
researchers\cite{vulnera1,vulnera2,vulnera3}. Whether nodes or links
should be considered as subjects of these failures depends on the
context: If the network under investigation is a model of the
Internet\cite{vulinf0}, a computer may break down unexpectedly,
which corresponds to a node failure. In the case of a transportation
network\cite{vulinf1}, a highway bridge may collapse after an
earthquake, in which case one rather speaks of a link failure. When
power grids are the subject\cite{vulinf2,vulinf3,vulinf4,vulinf5},
power lines may fail as well as the stations whence they emanate;
thus links and nodes can also fail simultaneously. Networks can also
suffer from intentional attacks as opposed to random failures.
Investigating the changes in the character and/or performance of
networks in any of these failure scenarios is crucial to
understanding the extent of damage they may suffer as well as
providing insight to how they can be reinforced and determining
which network types are more vulnerable.

The concept of vulnerability is associated with a system's ability
to fulfill its specific purpose under imposed conditions. The
purpose of the network dictates the conditions under which the
system is no longer functioning, thus it must be specified in order
to measure vulnerability. The primary purpose common for all
networks is transmission across links. Failure scenarios generally
involve a disconnection in the network, caused by a combination of
link failures, such that some source nodes can no longer send
signals or packages to some other target nodes. If a network is
divided into several components and its connectivity is impaired, it
is no longer capable of fully performing its transmission function.

A network may still function after the failure of certain nodes or
links. In the instance of a highway network, if the direct route
between two towns is no longer usable, the traffic may, if possible,
be directed to an alternative path. If there is such a path between
all pairs of nodes in the network after the failure events occur,
then the network is still connected. Even if there is no such path,
the traffic in the disconnected subnetworks do not have to come to a
halt. Thus the network is still somewhat able to fulfill its
function, although maybe not as efficiently and thoroughly as it
used to. Therefore, only focusing on maintaining the overall
connectivity is not always enough. Considering the case where the
failures push the network to the limit where it can barely stay
connected, the lack of alternative paths between node pairs inhibit
network performance. Controlling the connectivity is important, but
changes in network performance must also be monitored.

Network performance can be measured by various parameters which are
generally based on path measures such as shortest paths, random
walks, or some other measure between these two extremes. Such
performance measures often deal with how efficiently signals and
goods can be carried along the network. In an unrealistic model in
which adding links to a network has no cost, a network designer
would very likely want to connect each node to all others: In this
case any transmission would be carried out by moving along only one
link, and both the total number of alternative paths and the network
performance attain their maximum values. Hence complete networks
have the best performance and are least vulnerable, despite being
rare in real life. At the other extreme, a tree network is simply
connected but the failure of any single element partitions the
network and the performance is weak. Real life networks lie in
between these two examples, with varying degree distributions and
more complex connectivity patterns. When geographical constraints
exist, which is the case for any infrastructure network, adding
certain links can be costly or even impossible.

In an attempt to quantify vulnerability in this study, path measures
are used to rank the links in terms of their significance in
decreasing order, and then the failure of these links are considered.
Networks of equal sizes and average degrees but with different degree
distributions are tested, and a formal vulnerability measure is
devised based on some measured parameters. This vulnerability measure
is then used along with other local and global parameters in comparing
different network types. Our findings suggest that Erd\"{o}s-R\'{e}nyi
networks are efficient, and less vulnerable but lack a local structure
pattern, whereas scale free networks behave differently for varying
damage specifications. 

\section{FAILURE SIMULATIONS}
\subsection{Centrality Measures}

An important question to address when considering link failures is
how to rank the links in terms of their criticality. One extreme
would be to run the simulations for randomly failing links. A more
systematic approach would be to rank the links according to some
criteria. Over the years, network researchers introduced a number of
measures for ranking the elements of a network according to their
position and role in the topology. Such measures are generally
termed as centrality indices\cite{cent0,cent2,cent3}. One of
these indices, the shortest path betweenness centrality, is defined
to be the fraction of shortest paths between pairs of nodes in a
network that pass through an element. If there is more than one
shortest path between a given pair of nodes, then each such path is
given equal weight with the sum of the weights equal to one.
Shortest path betweenness can be thought of as a simulation in which 
a network with n nodes is considered and there are $n-1$ agents
at each node, each with the goal of reaching each one of the
remaining $n-1$ nodes using the shortest route possible. The
elements that have been visited the most have the highest
values.\cite{btwcent0,btwcent1}

Since this measure takes only the shortest paths into account, it
may lead to strong biases in certain situations. Consider a network
in which two clusters are connected only by two paths, one shorter
than the other. All the shortest paths between nodes of the two
different clusters will pass through the shorter of these paths, and
the longer path will therefore have a shortest path betweenness
value of zero. This longer path, however, is obviously not as
insignificant for the network as this measure suggests, since once
these two paths fail the network would be divided into two distinct
clusters that are not connected. As an alternative, random walks can
be considered instead of shortest paths. A simple random walk
suggests that a walker located at a specific node chooses to move
along on any one of the incident links with equal probability, and
continues moving until it finds itself at the target. Walks with
varying properties can be generated by manipulating the transition
probabilities\cite{rw}. The random walk betweenness of a link is
then defined as the number of times a random walk between node pairs
passes through that link, averaged over all node pairs
\cite{rwcent,btwrwcent}.

Another measure of the importance of a link to a network is the so
called average degree of a link, which is be calculated as the
average degree of the two nodes at the two ends of a link. Although
this measure is very blunt at distinguishing links connecting
low-degree nodes to high-degree nodes from those links that connect
two relatively average degree nodes, it can still be considered
useful because it is a very simple mechanism and far more easy to
calculate than the betweenness methods.

\subsection{Network Simulations}

Four types of networks with a fixed size (number of nodes $n=256$)
and average degree (average number of links per node $\bar{k}=8$) 
are investigated via simulations: Erd\"{o}s-R\'{e}nyi networks,
scale free networks, small world networks, and ring substrates.
The former two are created by first drawing a degree sequence
from the characteristic degree distribution: poisson distribution
for Erd\"{o}s-R\'{e}nyi and power law distribution for the scale 
free networks \cite{sf} (Obtaining this sequence is known to be
an issue for scale free networks\cite{graph4}). The process is
completed by randomly connecting nodes until the prescribed degree
sequence is reached \cite{graph0,graph2}. Small world networks are
produced using the Watts--Strogatz model, with a rewiring
probability that corresponds to a high clustering coefficient and
low shortest path length for this specific network size and average
degree\cite{sw0,sw1,sw2}. Once a network is formed, its links are
ranked according to shortest path betweenness, random walk
betweenness, average degree, and randomly. The highest ranked link
is then broken to start the simulation of a series of failures, and
this process is repeated until all links have failed.

The network reconfigures as failures occur: The degree distribution,
the paths, the betweenness values all change, and therefore the link
ranking also changes, sometimes drastically. Therefore recalculating
the link ranking after every failure is very crucial in the sense of
what the simulation physically represents.  As an example, consider
a highway network for which failure of a path is said to occur if
the average speed on it falls below a certain limit. Once this path
fails, drivers may choose move to an alternative path until this new
one becomes jammed too - and so the process evolves. Although
computationally cumbersome, the recalculation procedure is
significant in order to properly represent the effects of failures.

Depending on the chosen ranking method, the failure ratio and
the type of the network, the response in the network parameters
vary. In particular, five different parameters are monitored after
each link failure:

\begin{enumerate}
    \item {\textit{Ratio of Failed Links:}}
    The simulations move forward along a generated network by
failing links one at a time until the network becomes empty.
Therefore the independent variable in these simulations is the
number of failed links. For easier tractability, this number is
divided by the initial number of links in the network to obtain the
\textit{ratio of failed links}.
    \item{\textit{Fragmentation Ratio:}}
A simply connected network consists of nodes where any node can
reach any other. When disconnections occur and the network
partitions into components, the probability of reaching a target
node in one component becomes zero for the nodes in another
component. Therefore the number of disconnected components becomes a
reflective measure of how damaged or non-functional the network has
become. This number is normalized by the total number of nodes to be
fixed in the unit interval and called the \textit{fragmentation
ratio}. This parameter has been used in community detection
problems as well \cite{fragratio}. The sizes of the components is
also pertinent, and this issue has been been thoroughly investigated
\cite{discnp}. 
    \item{\textit{Ratio of Disconnected Node Pairs:}}
The disconnection of a network of $n=100$ into two components of 1
and 99 nodes or two components of 50 nodes each should not be
considered identical failures only because the number of
disconnected components are equal. Counting the number of node pairs that are disconnected
provides a simple way to quantify this measure. This number can be
normalized by the total number of node pairs to obtain the parameter
we call the \textit{ratio of disconnected node pairs}. Let $n_{i}$
denote the number of nodes in component $i$. Then,
\begin{equation}
\begin{array}{c} \text{\textit{ratio of}} \\ \text{\textit{disconnected}} \\ \text{\textit{node pairs}} \end{array} = \frac{\sum_{i\neq j}
n_{i}n_{j}}{n\left(n-1\right)/2}
\end{equation}
    \item{\textit{Clustering Coefficient:}}
The ratio of the number of links between the neighbors of a node to
the number of links between these neighbors if they were to form a
complete graph between them is called the \textit{clustering
coefficient}\cite{cc}. It is a measure of how locally redundant a
network is, since a node with a high clustering coefficient has a
high number of links connecting its neighbors, and the number of
alternating paths emanating from that node are also plenty.
Therefore this measure can also be interpreted as an indicator of
the abundance of alternative paths in the network.
    \item{\textit{Efficiency:}}
The sum of the inverses of the lengths of the shortest paths between
node pairs is the \textit{efficiency} of a network, a parameter in
the unit interval representing how short the shortest paths
are\cite{eff}. In mathematical form,
\begin{equation}
\text{\textit{efficiency}} = \sum_{i > j} \frac{1}{l_{ij}}
\end{equation}
where $l_{ij}$ denotes the length of the shortest path between nodes
$i$ and $j$.
\end{enumerate}

\section{SIMULATION RESULTS}

\begin{figure*}[!tbp]
        \subfloat{\includegraphics[height=4.4cm]{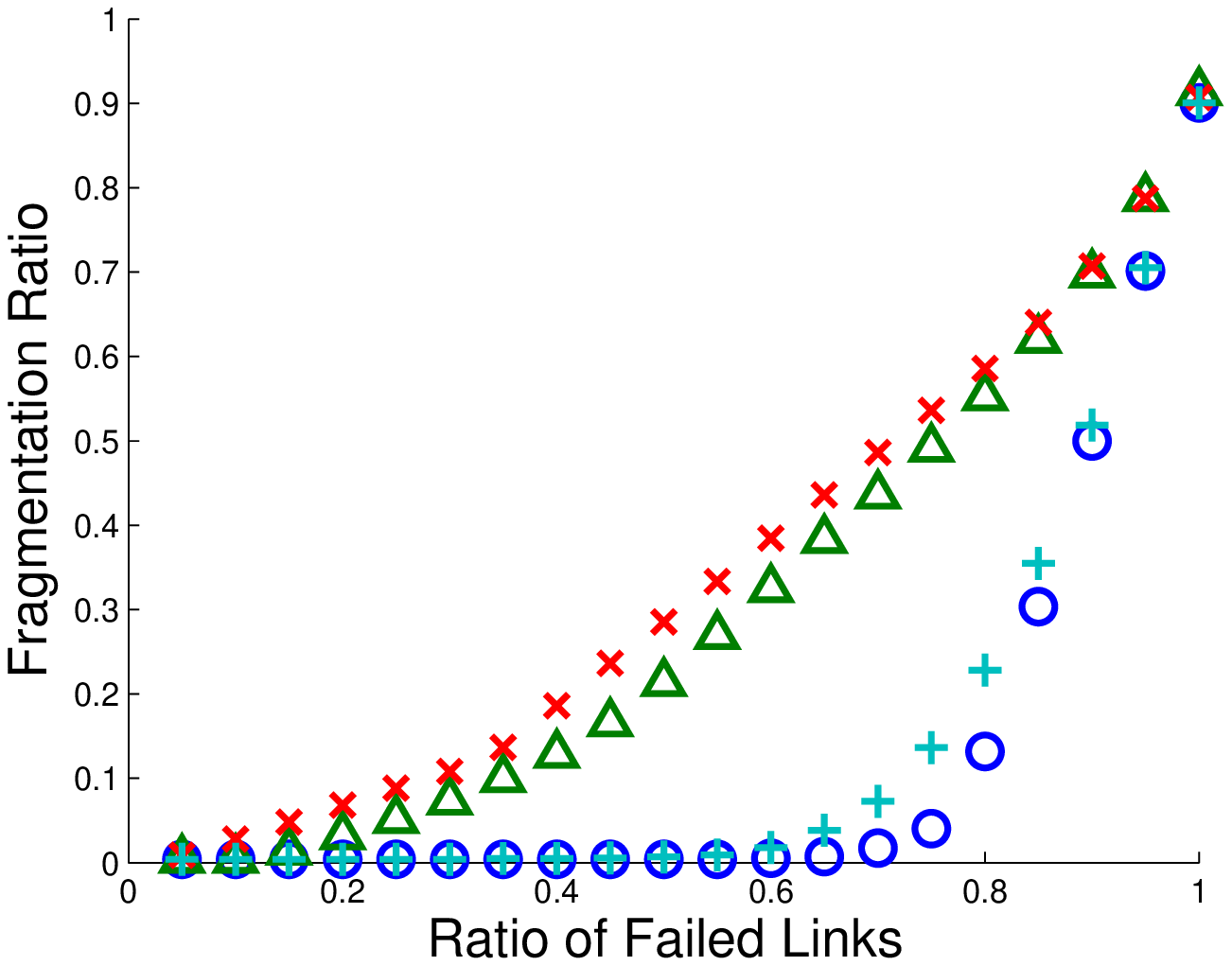}}
        \subfloat{\includegraphics[height=4.4cm]{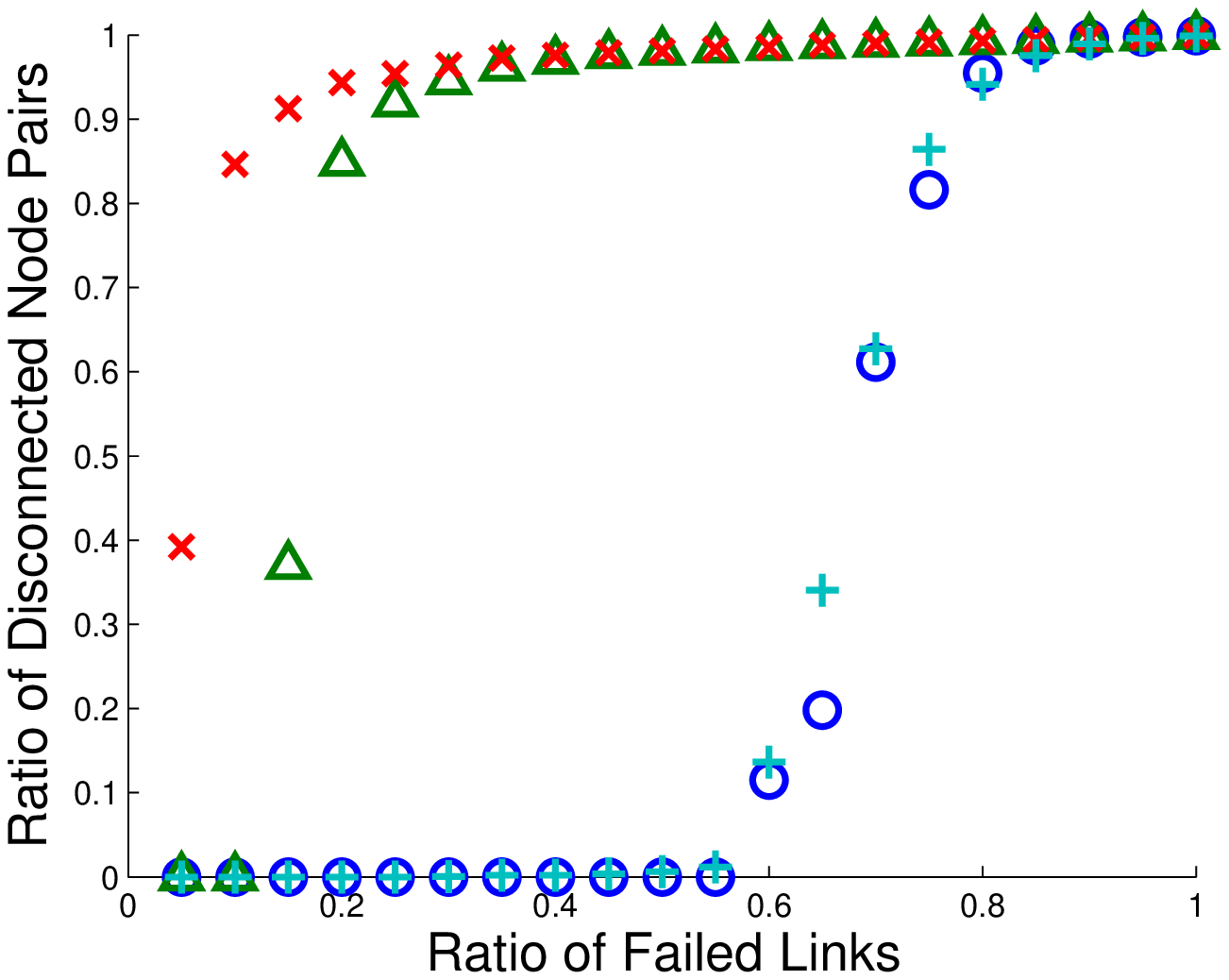}}
        \subfloat{\includegraphics[height=4.4cm]{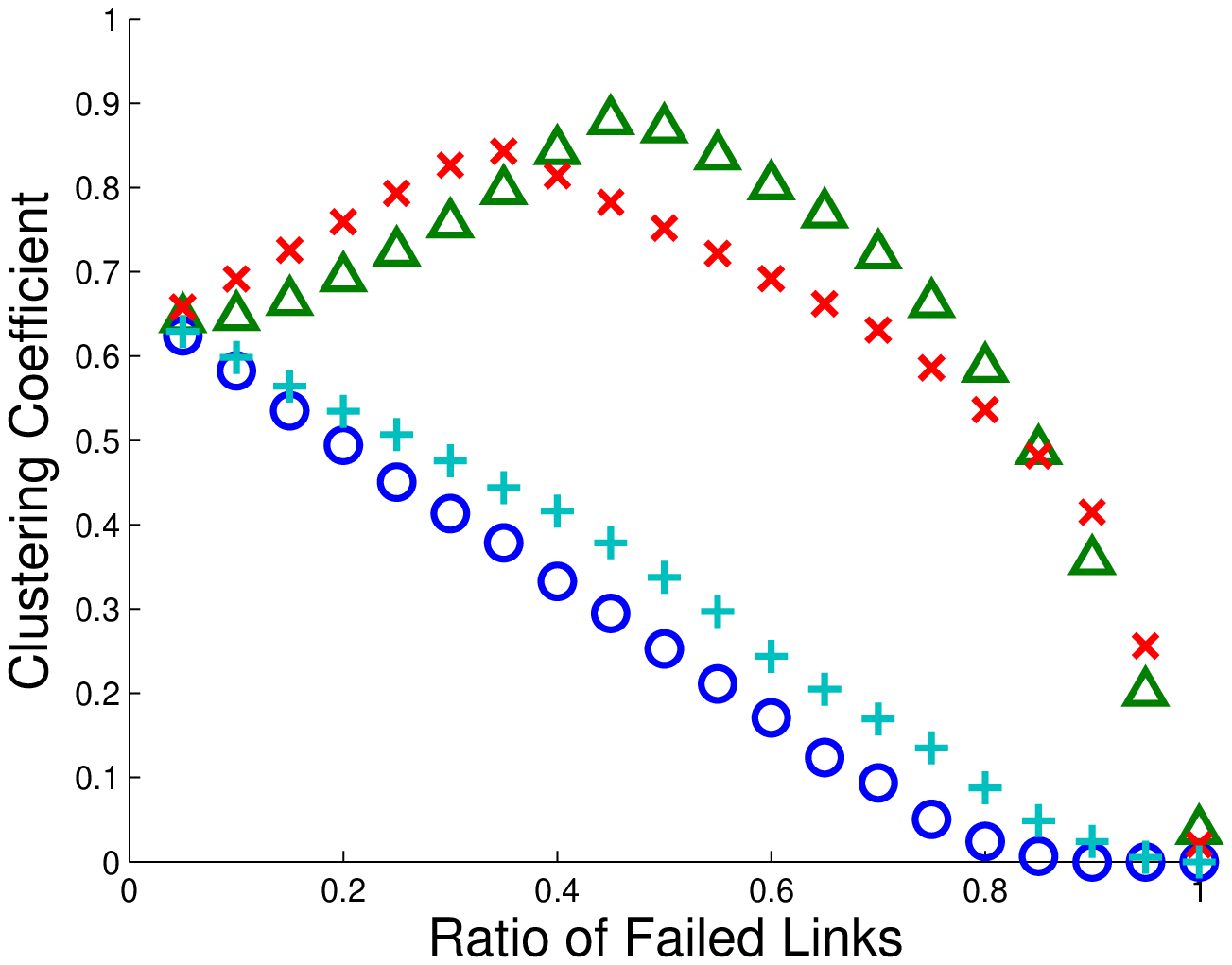}}
    \caption{The response of ring substrates of $n = 256$ and
    $\bar{k} = 8$, as the ratio of failed links is increased.
    Failure of the links with high betweenness values quickly
    disconnects the network into large components, each with
    relatively larger clustering coefficients. This process actually
    increases the clustering coefficient of the ring network,
    because the links with high betweenness values are also those
    that lie in regions of nodes with lower clustering coefficients,
    as can be seen in the third graph. The rapid increase in the
    ratio of disconnected node pairs happens simultaneously with the
    increase in the average clustering coefficient. ( X : random
    walk betweenness, $\Delta$ : shortest path betweenness, O :
    average degree , $+$ : random)} \label{fig_rs}
\end{figure*}

\begin{figure*}[!tbp]
        \subfloat{\includegraphics[height=4.4cm]{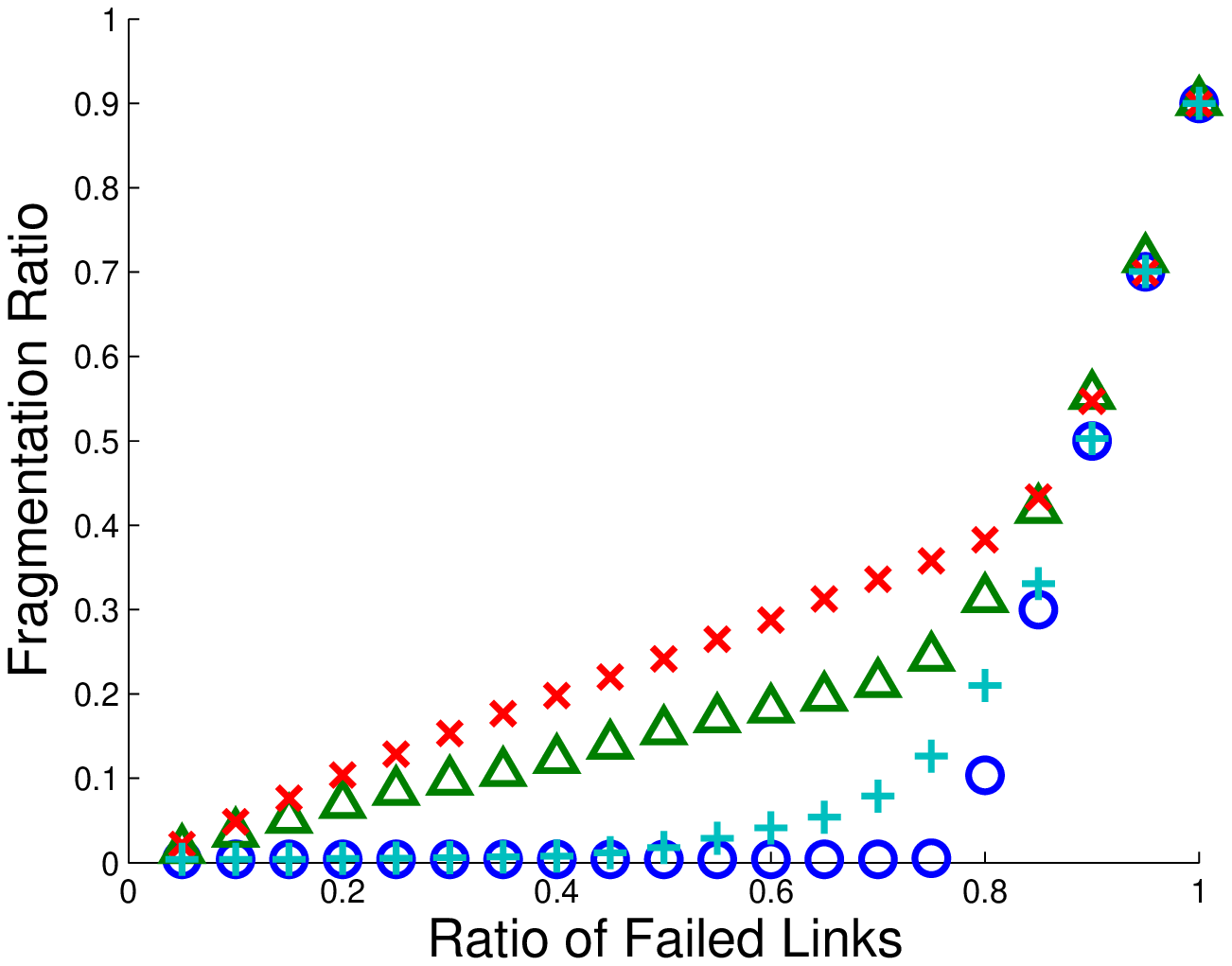}}
        \subfloat{\includegraphics[height=4.4cm]{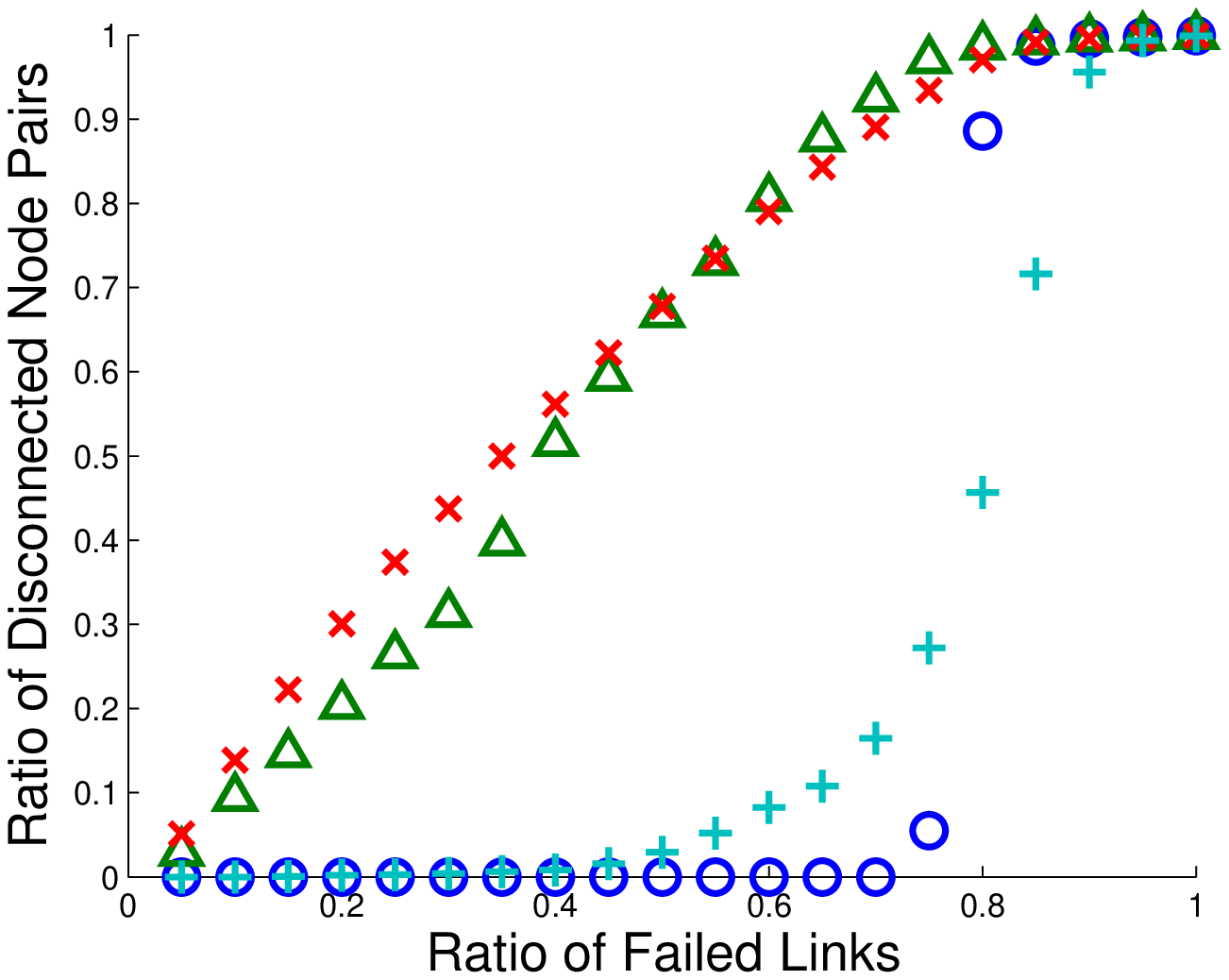}}
        \subfloat{\includegraphics[height=4.4cm]{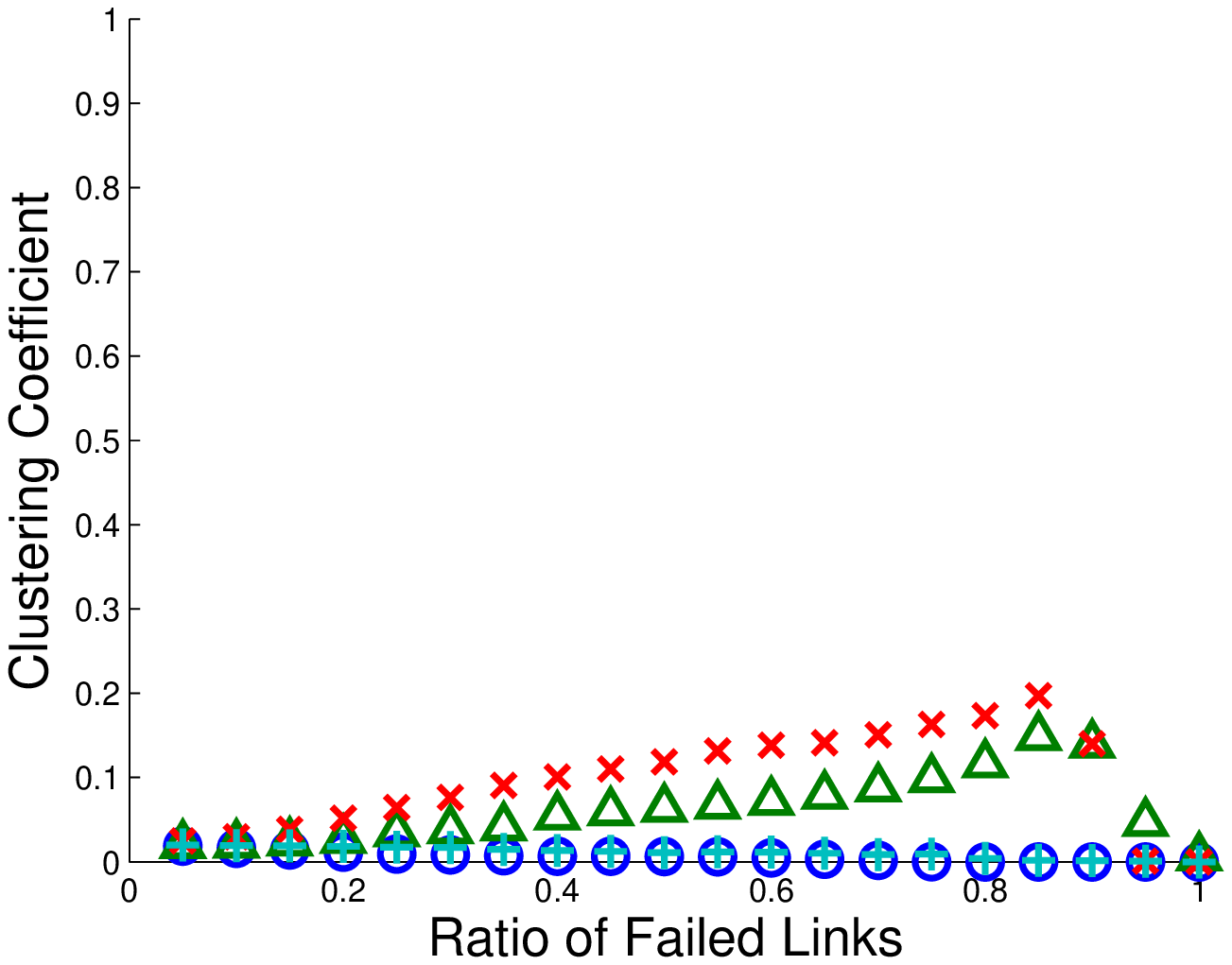}}
\caption{The response of Erd\"{o}s-R\'{e}nyi networks of $n = 256$
and $\bar{k} = 8$, as the ratio of failed links is increased. In
terms of fragmentation, ranking methods fail to differ by a good
margin. The fragmentation is almost bilinear with a slower initial
trend. Node-pair disconnection speed is very slow when compared to
ring substrates. The increase in the clustering coefficient is again
apparent for betweenness schemes. ( X : random walk betweenness,
$\Delta$ : shortest path betweenness, O : average degree , $+$ :
random)} \label{fig_er}
\end{figure*}

\begin{figure*}[!tbp]
        \subfloat{\includegraphics[height=4.4cm]{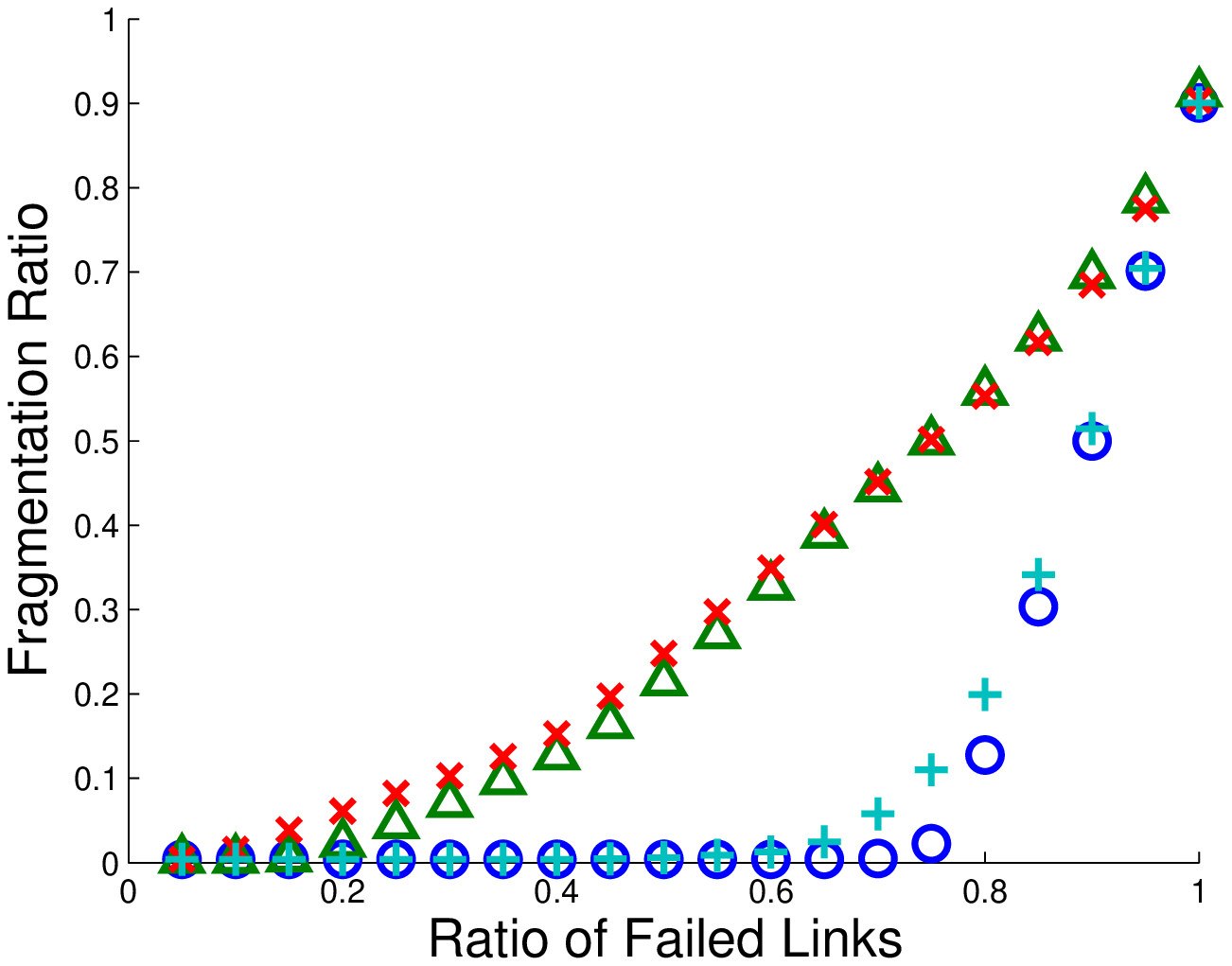}}
        \subfloat{\includegraphics[height=4.4cm]{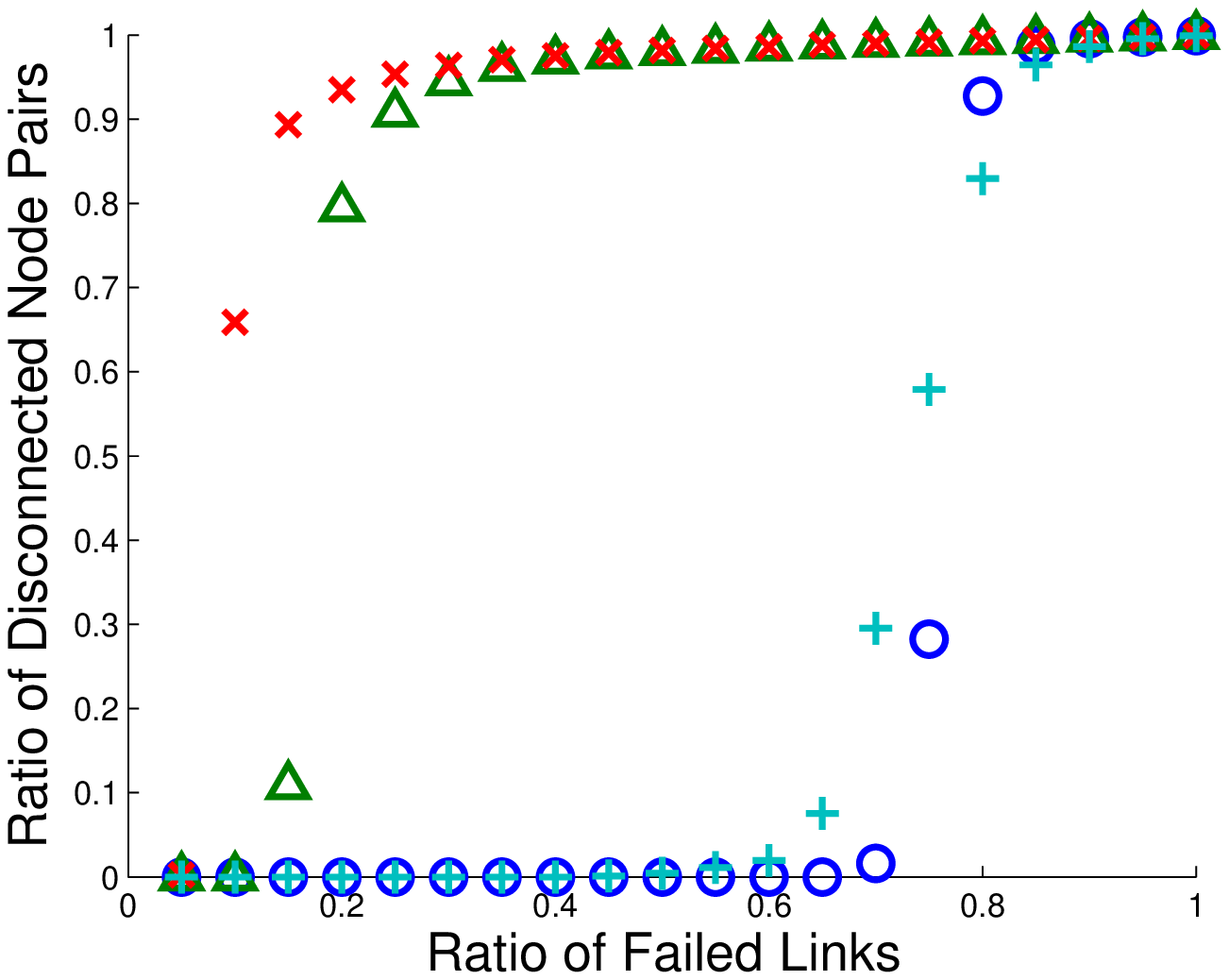}}
        \subfloat{\includegraphics[height=4.4cm]{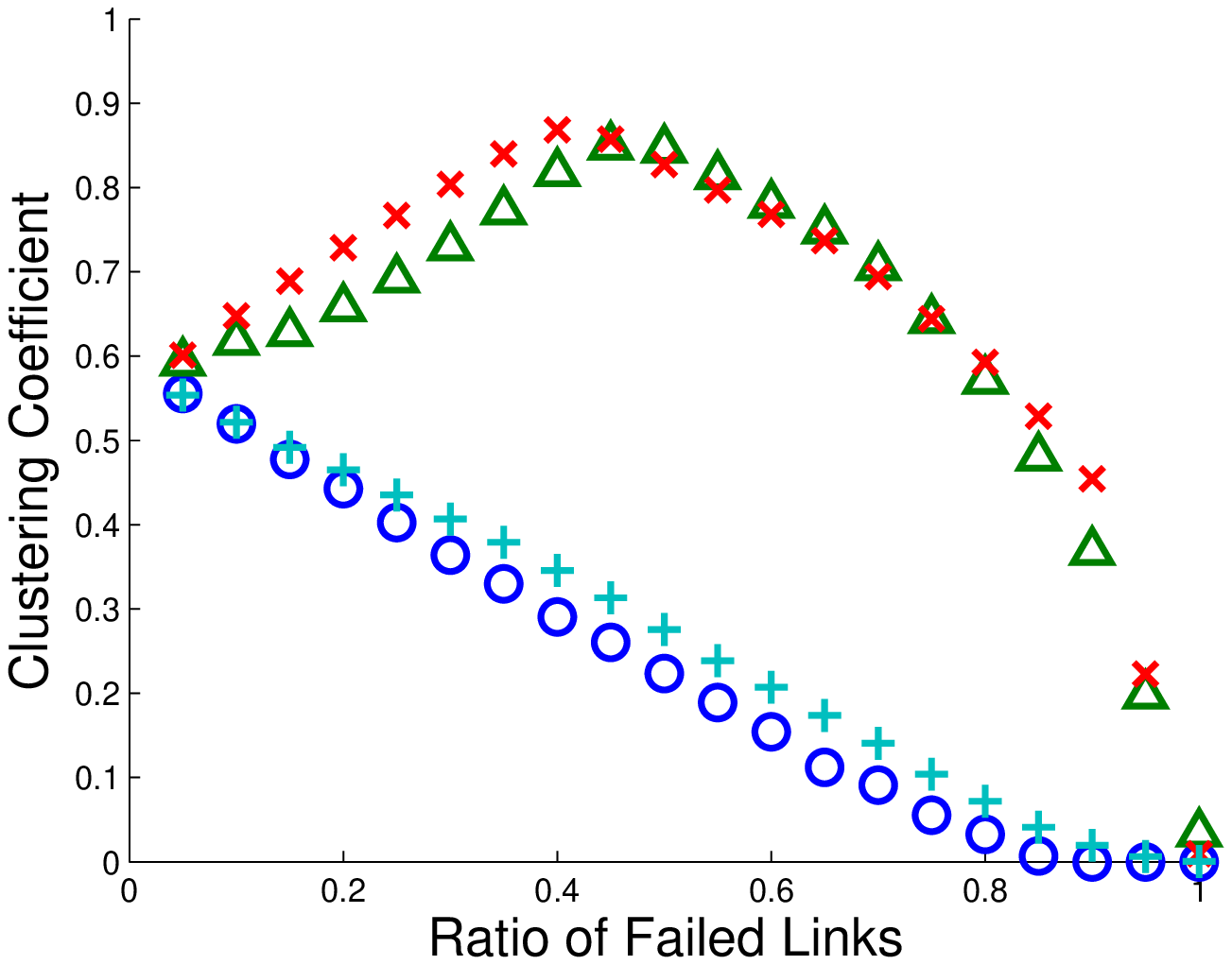}}

    \caption{The response of small world networks of $n = 256$ and
$\bar{k} = 8$ with rewiring probability $\beta=0.035$ \cite{sw0}, as the ratio
of failed links is increased. The results are almost identical to
that of the ring substrate. ( X : random walk betweenness, $\Delta$
: shortest path betweenness, O : average degree , $+$ : random)}
\label{fig_sw}
\end{figure*}

\begin{figure}[!tbp]
        \subfloat{\includegraphics[height=4.4cm]{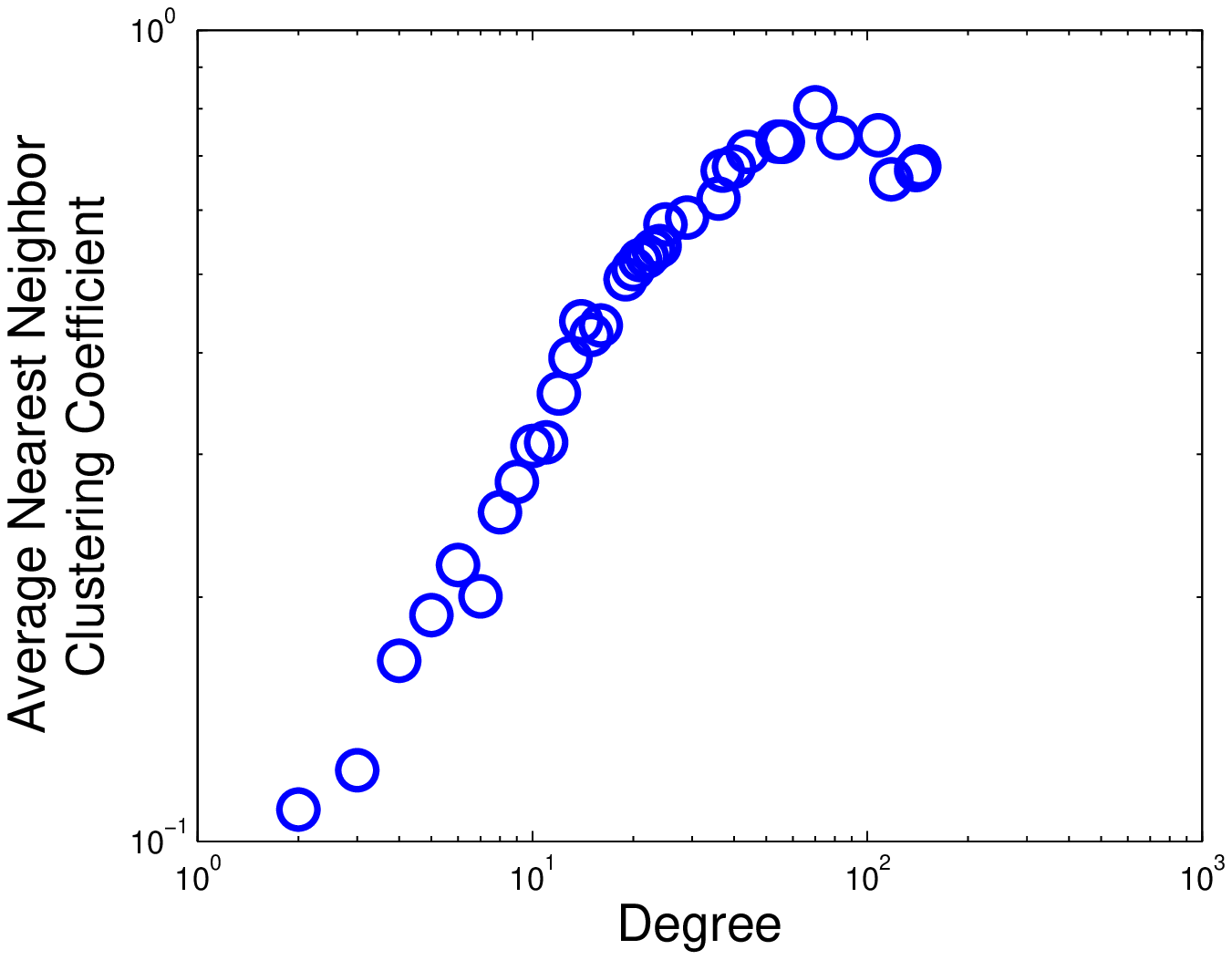}}
				\caption{The average nearest neighbor clustering coefficient as a function of the node degree for a typical scale free network with of $n = 256$ and $\bar{k} = 8$. The few high degree nodes have higher clustering coefficients compared to the low degree nodes.}
\label{avgcnn}
\end{figure}

The connectivity patterns of different network types are observed to
effect the simulation results, so we start this section by a brief
review of the network types considered. Ring substrates are ordered
networks with a degree distribution having zero variance. At the
other extreme, scale free networks have betweenness distributions
that obey the \textit{power law}. The average nearest neighbor
degree of scale free networks generally decrease as the node degree
increases, but the average nearest neighbor clustering coefficient
increases, as in Figure \ref{avgcnn} \cite{tez}. This observation is
not a contradiction and it in fact highlights the main connectivity pattern in a scale free
network: few hubs with low clustering coefficient values are connected
to many low-degree nodes that have higher clustering coefficients.
Erd\"{o}s-R\'{e}nyi networks lie somewhere between these two extremes,
however their totally random creation process inhibits clustering,
unlike the scale free networks where there are certain zones with high
clustering. Small world networks, formed by the random rewiring of a
few links of a ring substrate, are known to bear the relatively high
clustering of ring substrates, as well as the small shortest lengths
of Erd\"{o}s-R\'{e}nyi networks. In this sense small world networks
are an optimized version of ring substrates
\cite{sim0,sim1,sim2,sim3,sim4,sim5}.

The response of ring substrates to failures is shown in Figure
\ref{fig_rs}. The links that connect points which are farther apart
have higher betweenness values in these networks since they act as
shortcuts. When these long-range links fail, the extra stress in the
weakened zone is shared by the nearby elements. In other words, the
paths crossing that portion of the ring now have fewer alternatives,
and therefore the links that carry this extra traffic end up having
higher betweenness values. This damaged portion becomes the next
failure zone until the connection is no longer there, and the ring
becomes an arc. The fragmentation is almost linear for this failure
scheme, but the real damage can be seen in the ratio of disconnected
node pairs: before 20 percent of links fail, almost 95 percent of
node pairs are already disconnected. The random and average degree
failures are not extremely harmful in comparison, but networks
undergo a change reminding a phase transition around when 60 to 80
percent of links have failed for these schemes.

\begin{figure}[!tbp] \begin{tikzpicture} \GraphInit[vstyle=Normal]
\Vertex[x=1.5,y=2]{A} \Vertex[x=0,y=1]{C} \Vertex[x=0,y=3]{D}
\Vertex[x=3,y=2]{B} \Vertex[x=4.5,y=1]{E} \Vertex[x=4.5,y=3]{F}
\Edge(A)(B) \Edge(D)(C) \Edge(A)(C) \Edge(D)(A) \Edge(B)(F) \Edge(B)(E) \Edge(E)(F)
\end{tikzpicture}
\caption{A small graph illustrating the
bad neighbor concept. In this configuration, all nodes have
clustering coefficients equal to 1 except for $A$ and $B$ which have
values equal to $1/3$. Correspondingly, the link connecting $A$ and
$B$ has the highest betweenness value. The failure of this link increases
the clustering coefficients of these two nodes to 1 and keeps the others
unchanged while at the same time disconnecting the network. Thus the
clustering coefficient of the network increases, along with the
fragmentation ratio and the ratio of disconnected node pairs.
Nodes $A$ and $B$ are \textit{bad neighbors} of each other.} 
\label{bad_nbor}
\end{figure}
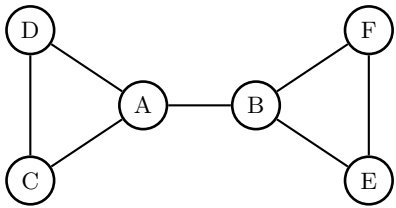

The increase  in the clustering coefficient observed for high
betweenness failures in Figures \ref{fig_rs} to
Figure \ref{fig_sw} may seem surprising in two aspects. The curves all
peak around a ratio of failed links of 0.5, but this is only a result
of the chosen network size and average degree. The more important point
is that the clustering coefficient is increasing, which is
counterintuitive but can simply be explained by the definition of
this parameter. If the number of neighbors of a node decreases, there
is a chance of an increase in the clustering coefficient of that node.
The process is depicted in Figure \ref{bad_nbor}. The failure of the
link between these nodes disconnects the network, but increases the
clustering coefficient of these networks to unity. In this case node
A is said to be a \textit{bad neighbor} of node B, and vice versa.
One other process that indirectly contributes to the increase of the
clustering coefficient is that a node with zero degree is no different
than a node with degree one in terms of its clustering coefficient,
hence the disconnection of this node leaves the average clustering
coefficient of the network unchanged. Considering these processes
in the larger scale, it may be concluded that the high betweenness
valued links must somehow be those links that contribute negatively
to the clustering coefficient, i.e. the bad neighbors. 

This result shows that betweenness and clustering coefficient are somehow
related. A low clustering coefficient signals the lack of alternative paths,
and in such cases some incident links become the only available route for a
node to connect with other regions. As a consequence, these links lying in
zones with low clustering coefficients have high betweenness values and
generally contribute to the bad neighbor effect: They inhibit clustering but
are also good transmitters.  Therefore it is reasonable to expect an increase
in the overall clustering coefficient when these links fail.

In Erd\"{o}s-R\'{e}nyi networks, the increase in the clustering
coefficient for the failure of links with high betweenness values is
smaller. The parameter peaks at the points where about 60 to 90
percent of all links have failed, which corresponds to a ratio of
disconnected pairs very close to 1. At this point
the network consists of tiny connected subgraphs that have the
smallest number of bad neighbors possible, which causes the slight
increase in the clustering coefficient. It should be noted that this
increase can also be caused by the very low clustering values at the
initial formation.

\begin{figure*}[t]
        \subfloat{\includegraphics[height=4.4cm]{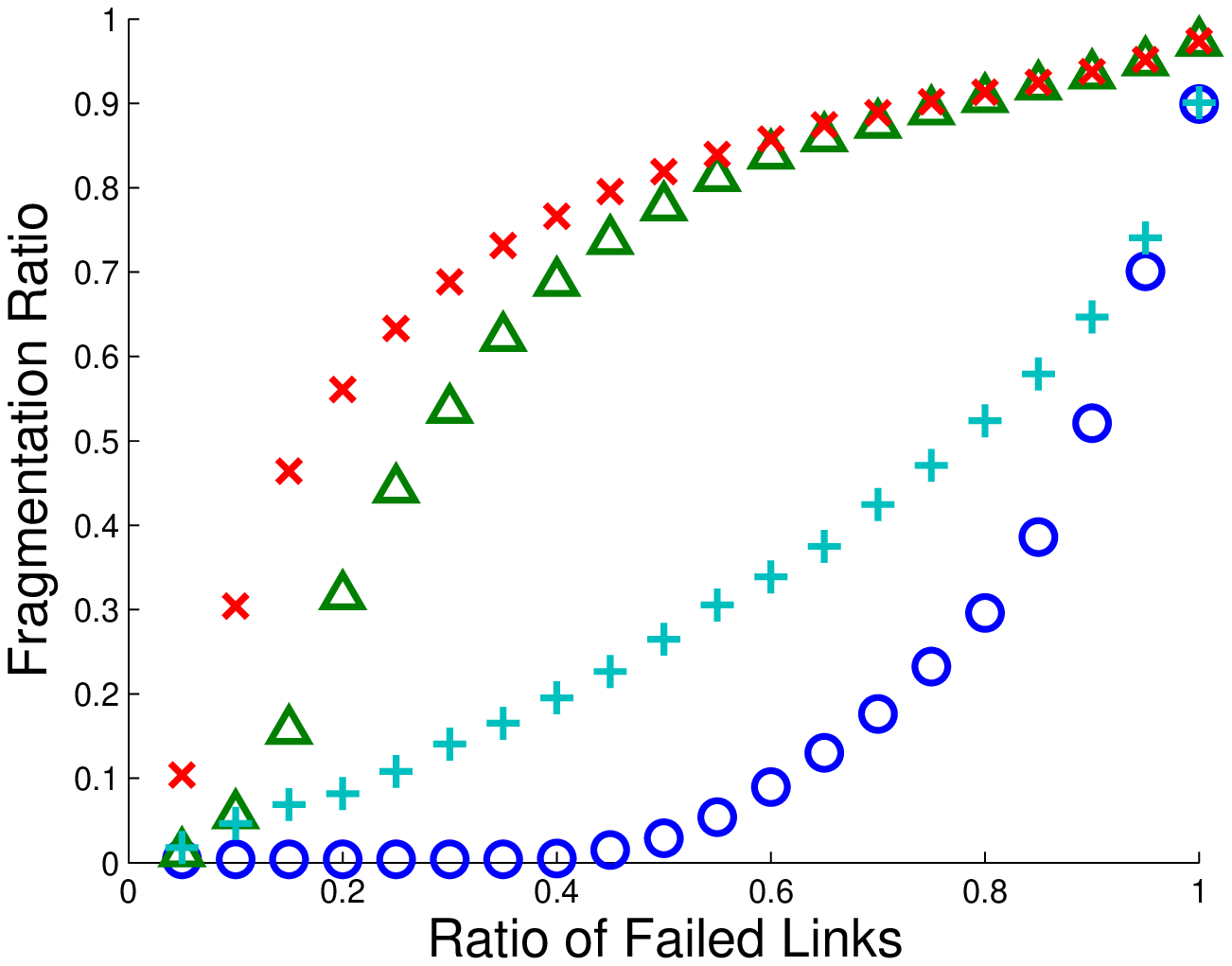}}
        \subfloat{\includegraphics[height=4.4cm]{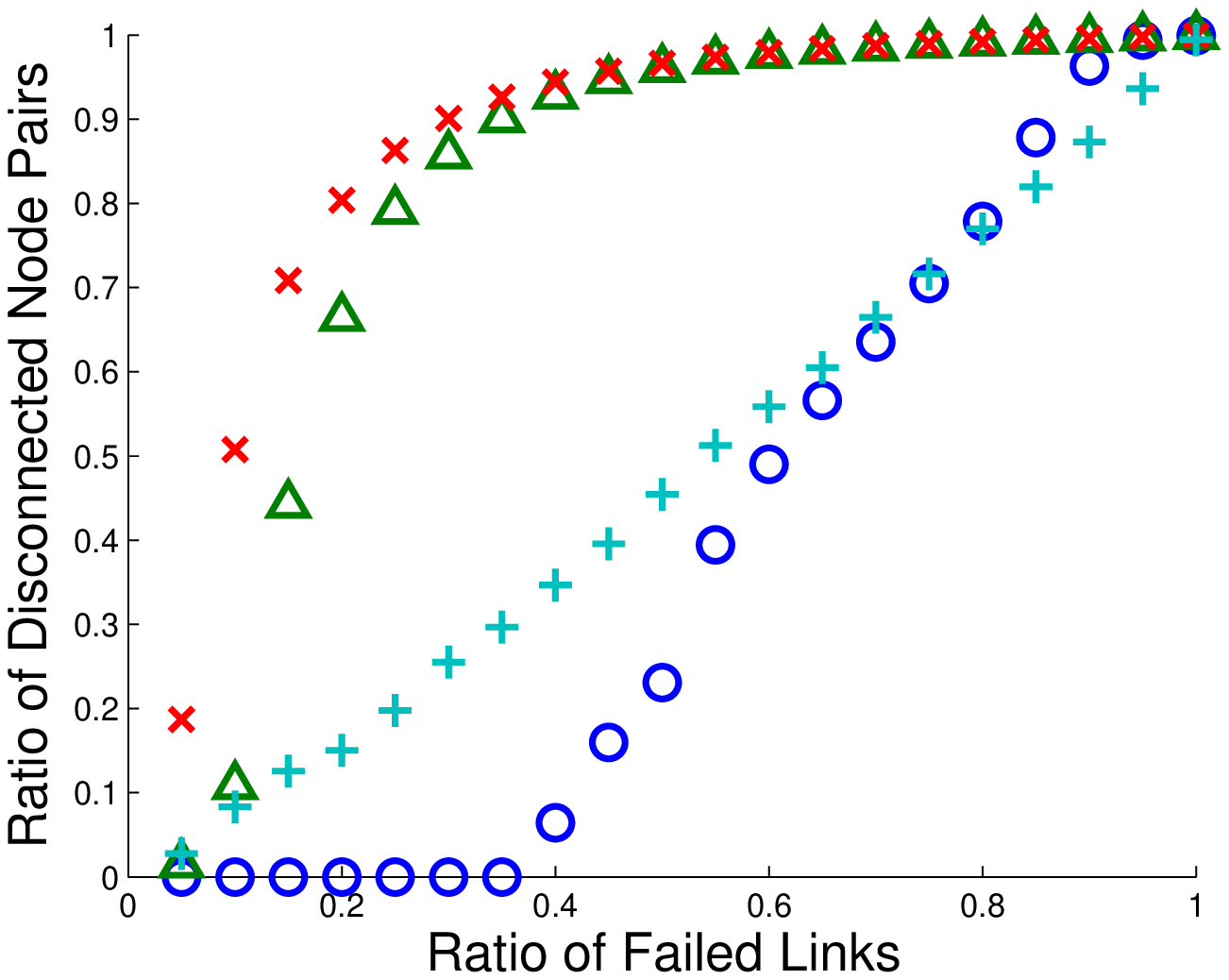}}
				\subfloat{\includegraphics[height=4.4cm]{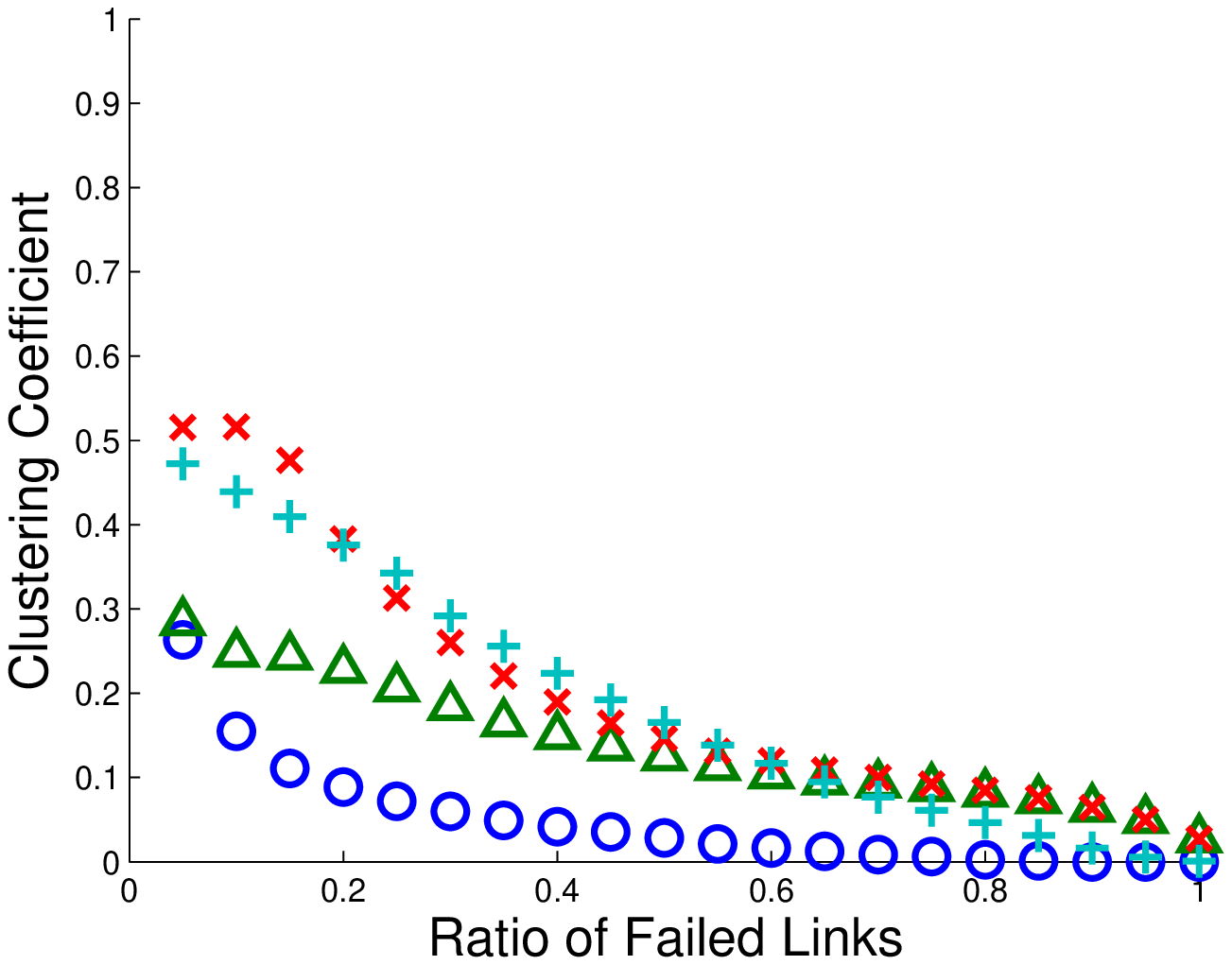}}
\caption{The response of scale free networks of $n = 256$ and
$\bar{k} = 8$, as the ratio of failed links is increased. The
fragmentation is almost bilinear, but with a very steep initial
trend, and node pair disconnections quickly saturate. There is no
increase in the clustering coefficient.  ( X : random walk
betweenness, $\Delta$ : shortest path betweenness, O : average
degree , $+$ : random)} \label{fig_sf}
\end{figure*}

Figure \ref{fig_sw} shows that small world networks, differing only
by a small percentage of link connections from ring substrates, fail
to respond very differently than their ancestors. Comparing
these two networks in the smaller failure zones (for example only up
to the rewiring probability) rather than in the whole unit interval
might be more appropriate. However the parameters used to measure the network
response are not sensitive enough for small failures, thus making this
comparison is not rewarding.

Scale free networks seem to be most prone to fragmentation, as can
be observed in Figure \ref{fig_sf}. Interestingly, they do not
demonstrate the increase in the clustering coefficient like other
networks. On the contrary, random and average degree schemes exhibit
a very fast initial decrease. This different behaviour exhibited by
scale free networks is caused by the exceptional betweenness
distribution of these networks: Few critical links have
exceptionally high betweenness values, and many have low betweenness
values. As those few critical links fail (the 0 to 10 percent of
failed links zone), the clustering coefficient seems to be
non-decreasing for betweenness schemes. At the end of these failures,
both the fragmentation ratio and the ratio of disconnected node
pairs are very high, which suggests that the network then consists
of several disconnected components with similar sizes (note how low
variance of component sizes result in higher ratio of disconnected
node pairs). The variance of the betweenness distribution quickly
decreases such that the remaining formation does not allow any link
to stand out as a high betweenness element. Failure of links with
high betweenness values increase the clustering coefficient;
conversely, failure of low betweenness valued links tend to decrease
the clustering coefficient. So once a scale free network has lost
its few high betweenness links, the bad neighbor effect is no longer
pertinent.

It can be observed in all cases that betweenness methods are far
more effective in breaking down a network when compared to the
average degree and random failure schemes. The random walk
betweenness is generally more dangerous than the shortest path
betweenness, as observed in Figures \ref{fig_rs} to \ref{fig_sf}. However the time required to compute the
random walk betweenness repeatedly for large networks can be
inhibiting, and the shortest path betweenness may provide a very
practical estimate of random walk betweenness that is far more easy
to calculate. Another observation for all network types is that the
average degree failure scheme appears to be less destructive than
random failures. This is actually to be expected since this scheme
especially protects the low degree nodes. High average degree links
are those that connect high degree nodes; these nodes have many
other alternative paths, and even if a few of their incident links
fail, the remaining ones may suffice to maintain performance. It
therefore stands to reason that the probability of disconnection in
the average degree scheme should be smaller than that for random
failures.

\begin{figure*}[!tbp]
        \subfloat{\includegraphics[height=4.4cm]{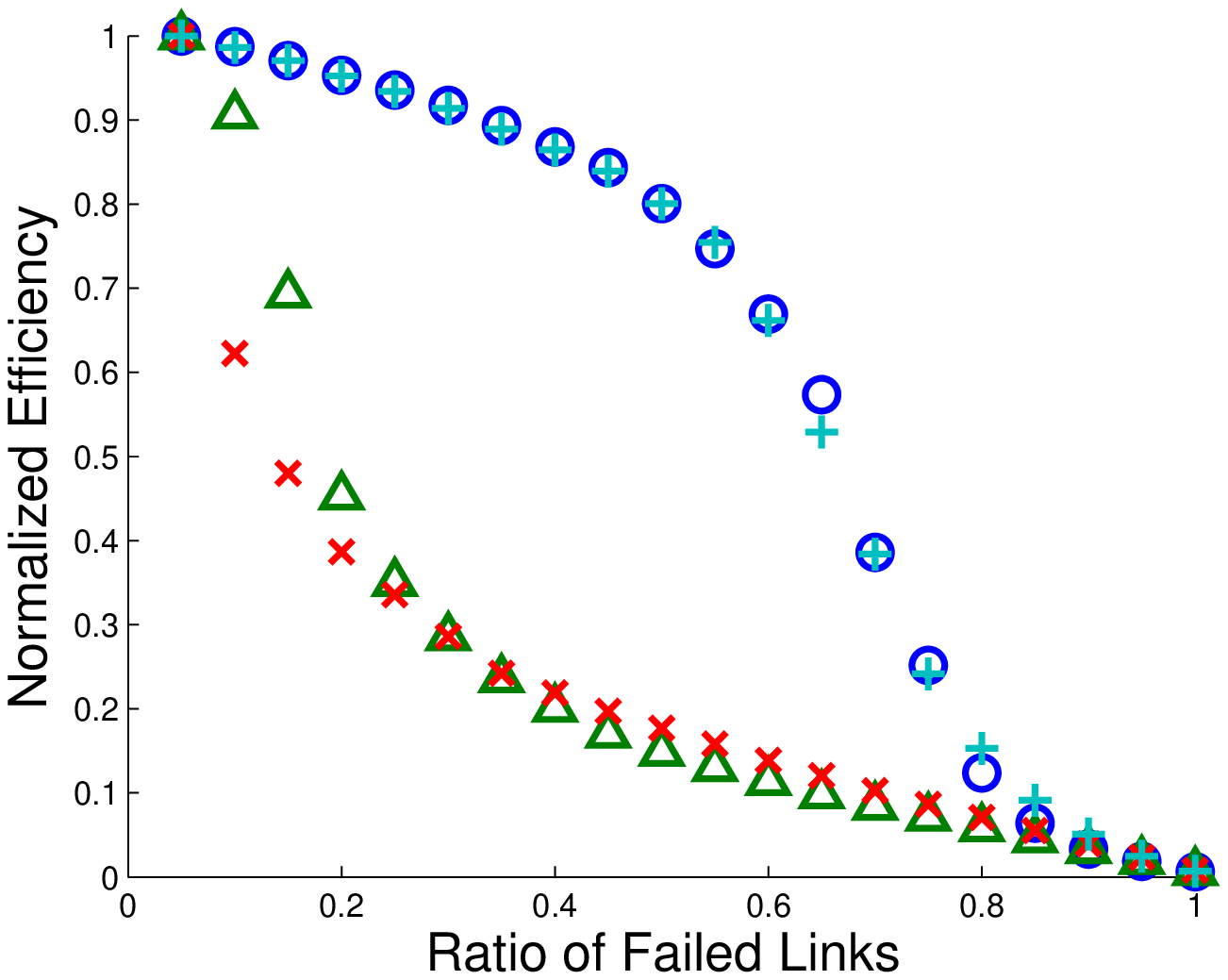}}
        \subfloat{\includegraphics[height=4.4cm]{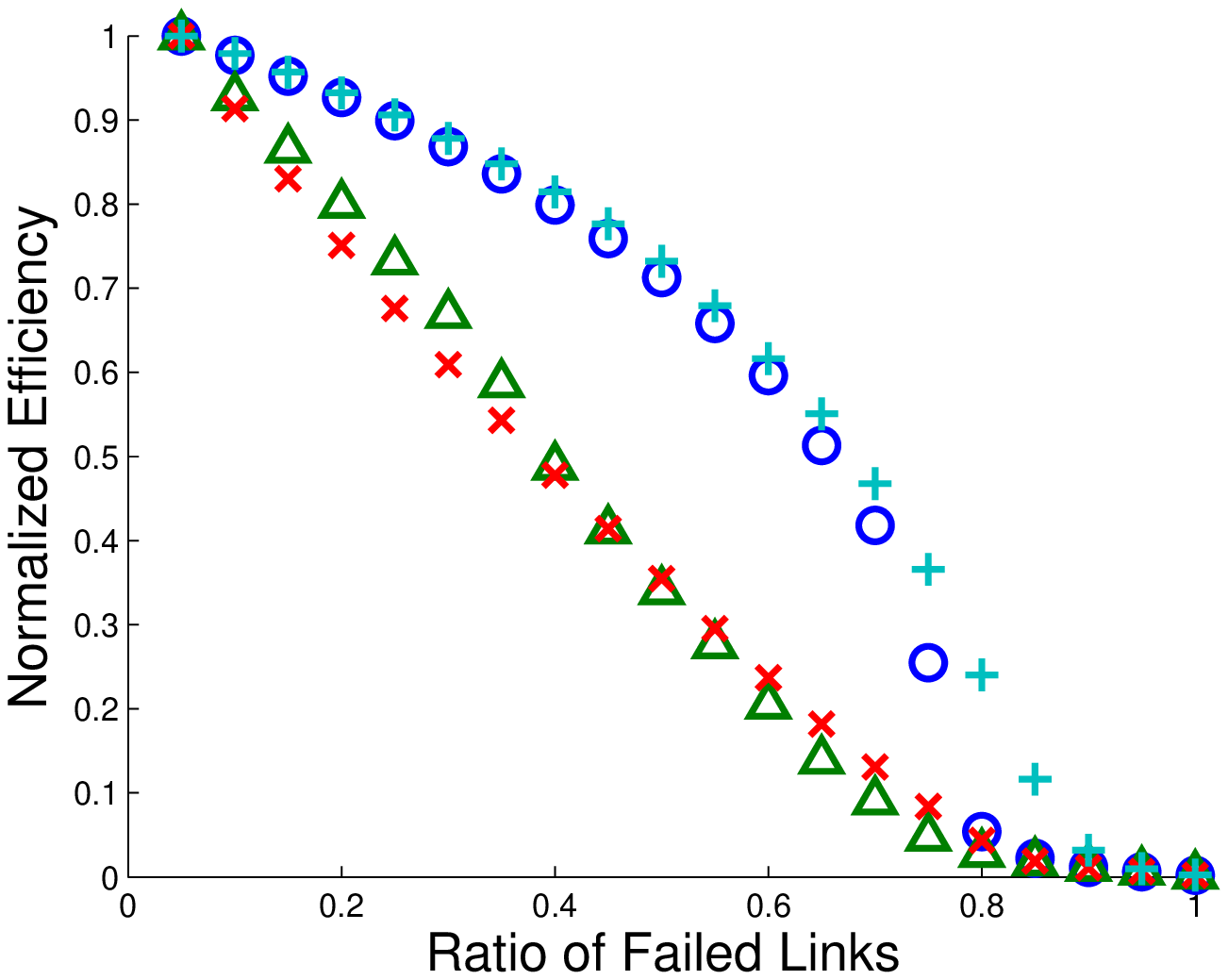}} \\
        \subfloat{\includegraphics[height=4.4cm]{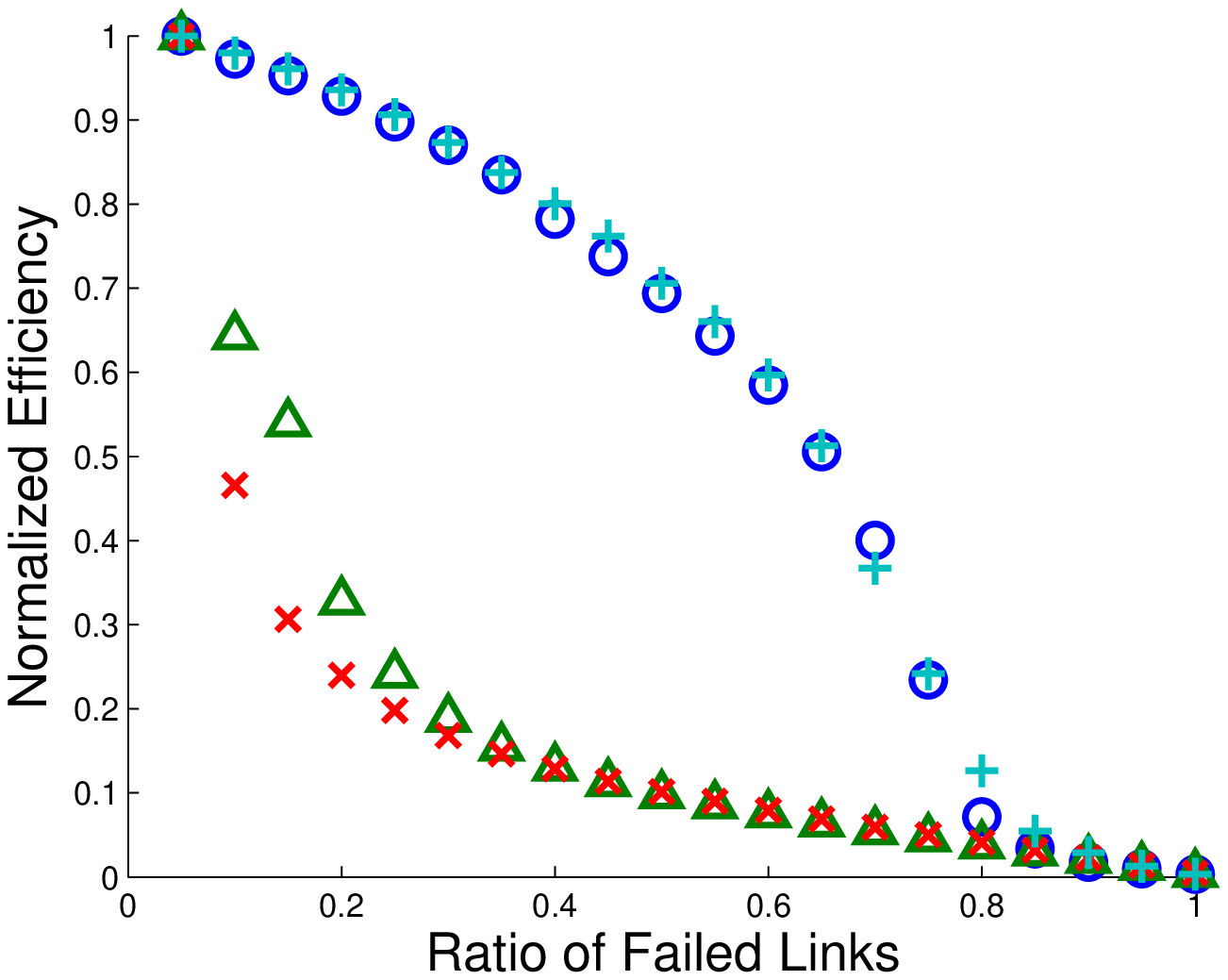}}
        \subfloat{\includegraphics[height=4.4cm]{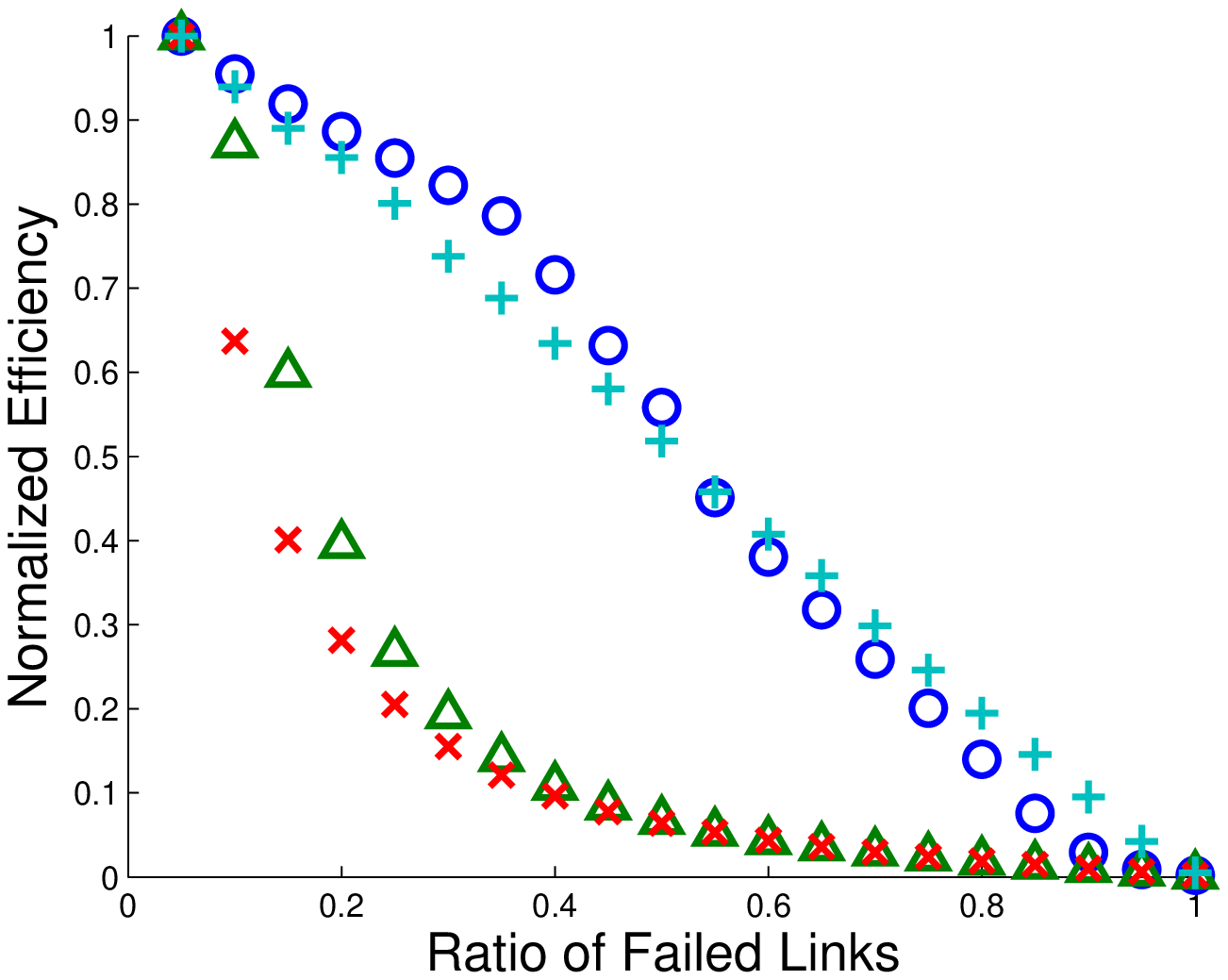}}
\caption{The efficiency response for ring substrates (upper left),
Erd\"{o}s-R\'{e}nyi (upper right), small world (lower left) and
scale free networks (lower right) of $n = 256$ and $\bar{k} = 8$,
respectively. ( X : random walk betweenness, $\Delta$ : shortest
path betweenness, O : average degree , $+$ : random)}
\label{fig_eff}
\end{figure*}

Since efficiency is strongly correlated with the ratio of
disconnected node pairs, the responses of the networks to failures
in terms of efficiency seem generally in line with what one might
anticipate. Figure \ref{fig_eff} reveals the efficiency response,
normalized by the efficiency value of each network at its initial
state. Scale free networks undergo a rapid efficiency decrease for
average degree and random failure schemes, unlike other network types.
For betweenness failures, Erd\"{o}s-R\'{e}nyi networks respond strongly,
the decrease behaves linearly unlike the exponential decrease in the
 other network types.

\section{VULNERABILITY}

The general formalism for the reliability or the vulnerability of
networks have so far been associated with primal cut
sets\cite{rel0,rel1,rel2}. In this section, we discuss the use of a
different approach to quantify what may be termed to be the
$vulnerability$ of a network.

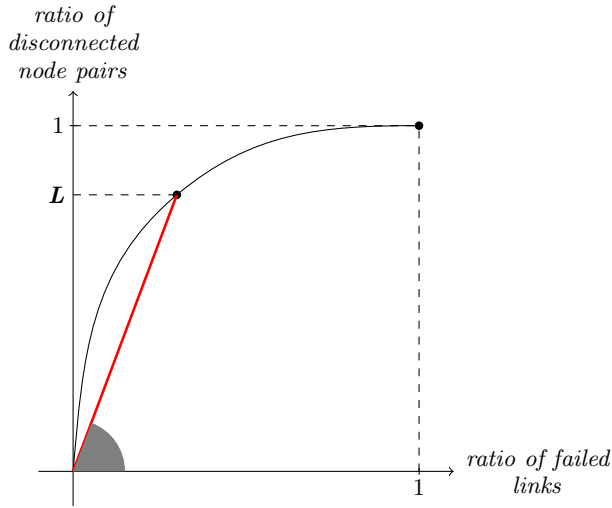
\begin{figure}[!tbp]
\begin{tikzpicture}[domain=0:1,scale=4.6]
\draw[->] (-0.1,0) -- (1.1,0) node[right,text width=2cm,text badly centered] {\textit{ratio of failed links}};
\draw[->] (0,-0.1) -- (0,1.1) node[above,text width=2.2cm,text badly centered] {\textit{ratio of disconnected node pairs}};
\coordinate [label=left: \textbf{\textit{L}},text width=1.5cm,text badly centered] (xint) at (0, 0.8);
\coordinate [label=below:$1$] (xbound) at (1,0); 
\coordinate [label=left:$1$] (ybound) at (0,1);
\draw (1,0.01)--(1,-0.01);
\draw (0.01,1)--(-0.01,1);
\coordinate (P) at (0.3, 0.8); \fill (P) circle (0.35pt);
\coordinate (end) at (1,1); \fill (end) circle (0.35pt);
\draw (0,0) to [out=85,in=220] (P);
\draw (P) to [out=40,in=180] (1,1);
\draw[color=red,line width=1pt] (0,0)--(P);
\draw[dashed] (xint)--(P);
\draw[dashed] (xbound)--(end);
\draw[dashed] (ybound)--(end);
\fill[color=gray] (0.15,0) arc (0:69:0.15cm) -- (0,0);
\end{tikzpicture}
\caption{A graph revealing the calculation of the vulnerability measure. The black curve depicts the change in the ratio of disconnected node pairs as the ratio of failed links increase, for the shortest path betweenness failure scheme. The red line is the linear approximation of this arbitrary curve, drawn from the origin to the black curve at the saturation level $L$. The slope of this red line, or the highlighted angle, is the vulnerability of the network.}
\label{fig_vuln}
\end{figure}

\begin{figure*}
\begin{tikzpicture}
\node[draw,circle,minimum size=2.5cm,fill=blue!25] at (1.25,0) {$n_{1}=244$};
\node[draw,circle,minimum size=0.4cm,fill=blue!25,label={below:$n_{2}=12$}] at (3.25,0) {};
\node[draw,circle,minimum size=1.25cm,fill=blue!25] at (6.75,0) {$n_{1}=128$};
\node[draw,circle,minimum size=1.25cm,fill=blue!25] at (8.75,0) {$n_{2}=128$};
\node[draw,circle,minimum size=0.75cm,fill=blue!25] at (12,0.6) {};
\node[draw,circle,minimum size=0.75cm,fill=blue!25] at (12,-0.6) {};
\node[draw,circle,minimum size=0.75cm,fill=blue!25] at (13,0.8) {};
\node[draw,circle,minimum size=0.75cm,fill=blue!25] at (13,-0.8) {};
\node[draw,circle,minimum size=0.75cm,fill=blue!25] at (14,0.8) {};
\node[draw,circle,minimum size=0.75cm,fill=blue!25] at (14,-0.8) {};
\node[draw,circle,minimum size=0.75cm,fill=blue!25] at (15,0.6) {};
\node[draw,circle,minimum size=0.75cm,fill=blue!25] at (15,-0.6) {};
\node at (13.5,0) {$n_{1\ldots8}=32$};
\end{tikzpicture} 
\caption{ Three disconnection cases for networks with $n = 256$ that approximately correspond to the three saturation levels used in this study, 0.1, 0.5 and 0.9, respectively. The disconnection of a relatively small group of nodes from the giant component depicts a saturation level of 0.1, whereas more than several components with equal sizes indicate a level of 0.9.}
    \label{satur}
\end{figure*}
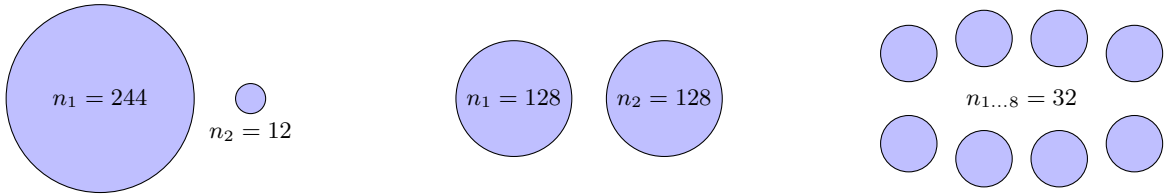

The results of the simulations indicate that
betweenness methods are very efficient in determining the critical
links in a network, and that their failures cause quick
fragmentation. Therefore considering this response of a network
against the failure of its most central links in order to formalize
a vulnerability measure would be reasonable. It has been
 stated that shortest path betweenness is a good estimate
for modeling the most detrimental link failures. Therefore here we
choose to apply the shortest path betweenness as the failure regime
in our calculations of vulnerability. In this context, we define the
vulnerability of a network as a measure of the average speed by
which the decomposition of the network causes a specified ratio of
node pairs to become disconnected. One must subjectively decide on
this ratio of disconnected node pairs above which the network is
accepted to be performing inadequately. If this ratio, which we
will call \textit{the saturation level}, is $L$, then the
vulnerability corresponding to this saturation level is defined as:

\begin{equation} vulnerability = \frac{L}{\text{ratio of failed links at $L$}} . \end{equation}

This definiton of vulnerability can also be interpreted as the
average percent of node pairs that are disconnected per unit
percentage link failures below the given saturation level, 
as shown in Figure \ref{fig_vuln}.

\begin{figure*}[!tbp]
        \subfloat{\includegraphics[height=4.4cm]{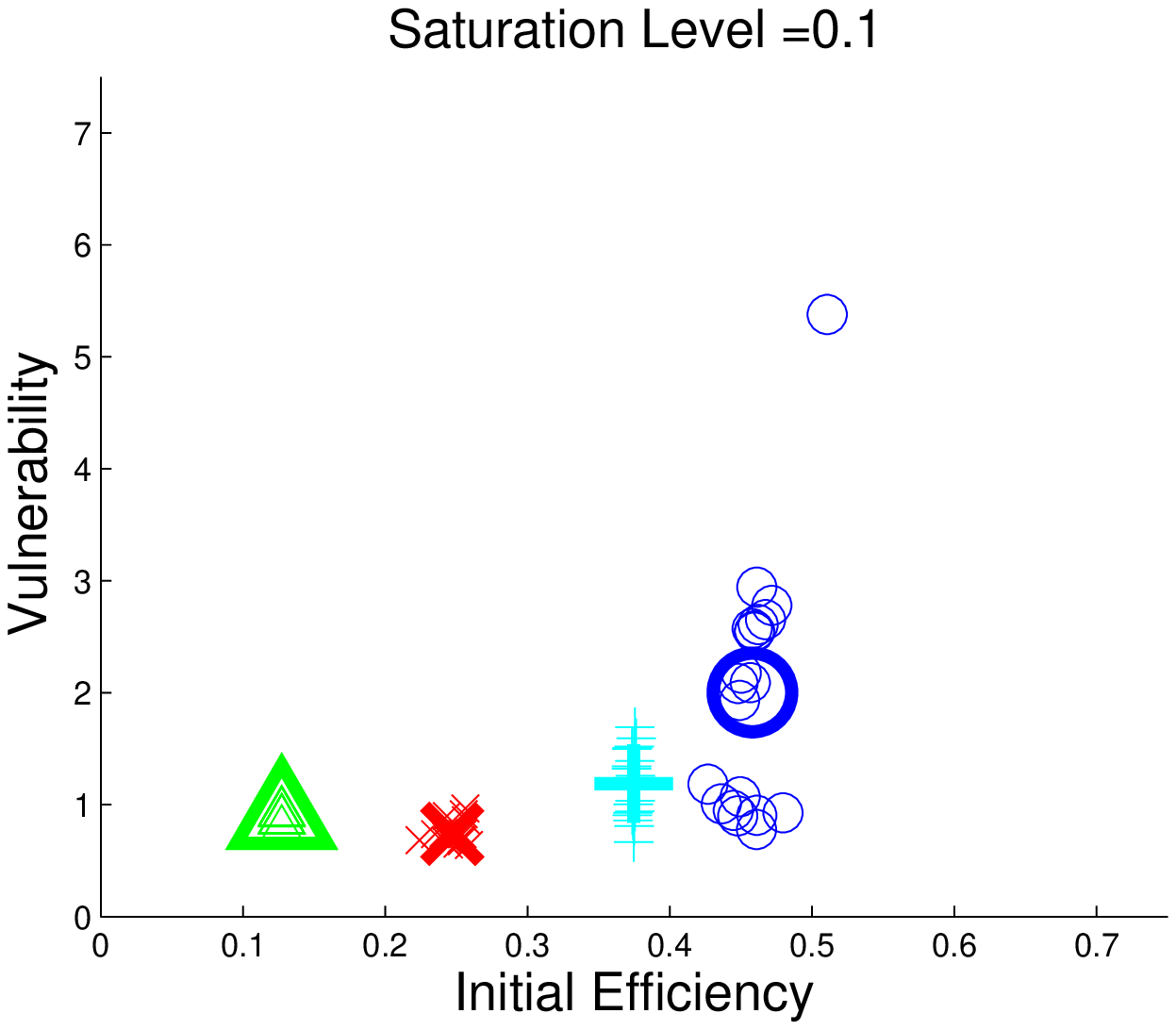}}
        \subfloat{\includegraphics[height=4.4cm]{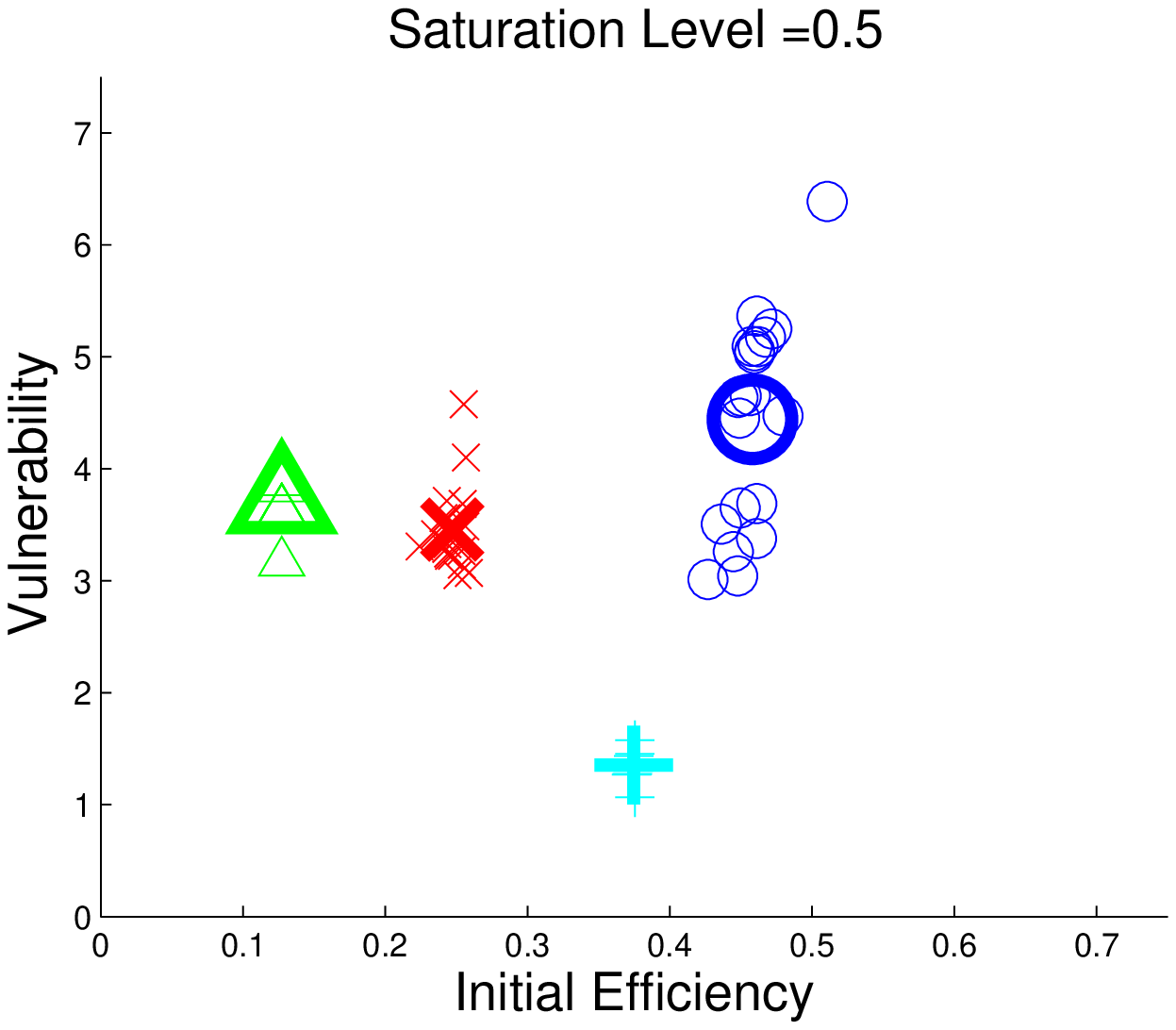}}
        \subfloat{\includegraphics[height=4.4cm]{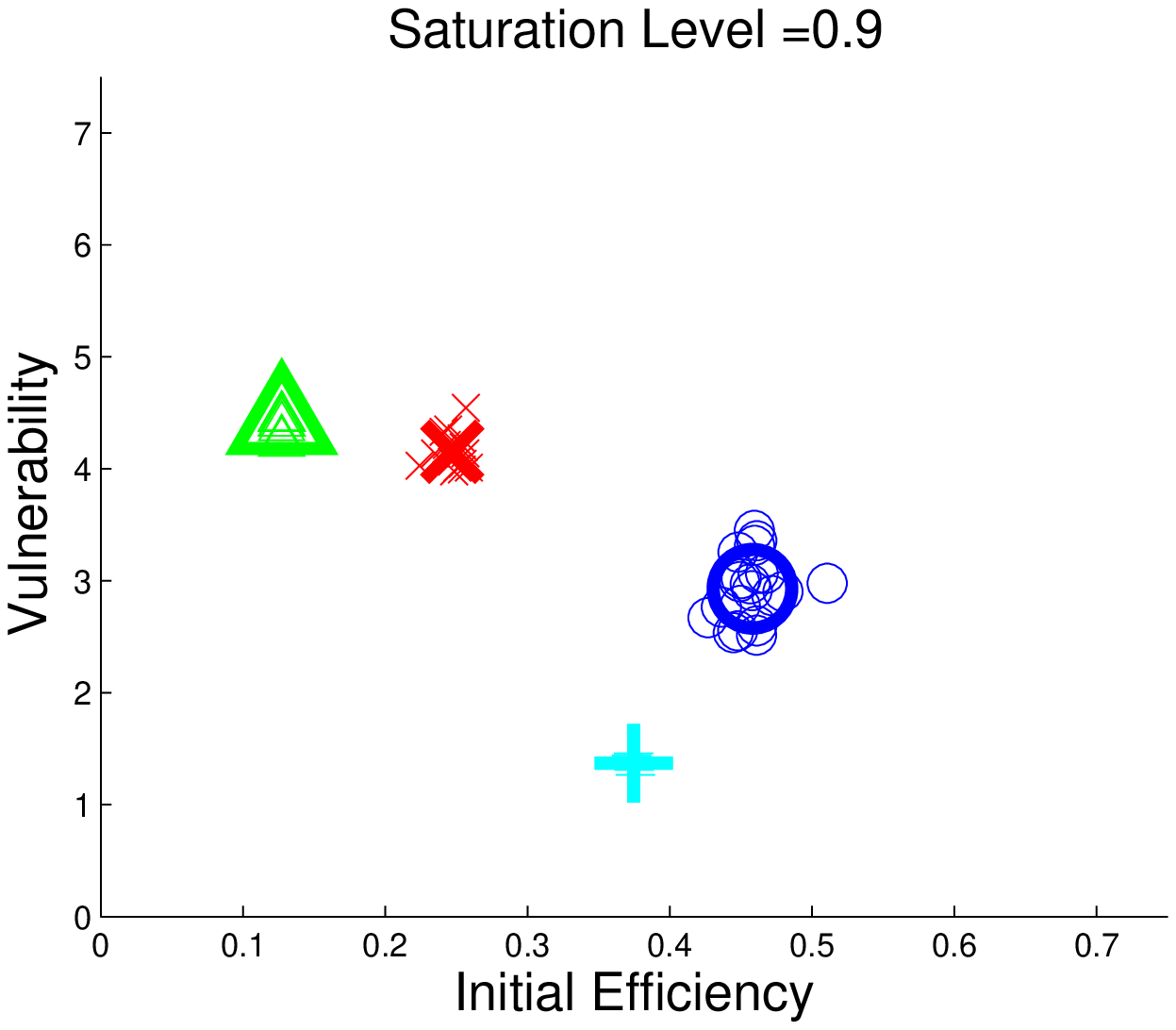}}

    \caption{ Vulnerability versus initial efficiency for the
    four network types and saturation levels of 0.1, 0.5 and 0.9.
    Smaller markers depict results of a single network realization,
    and the bigger markers represent the mean points of such
    realizations. For lower saturation levels scale free networks are
    more vulnerable but also more efficient. Ring substrates, small
    world networks and  Erd\"{o}s-R\'{e}nyi networks differ only slightly
    in terms of their vulnerabilities. For increased
    saturation levels of 0.5 and 0.9,  Erd\"{o}s-R\'{e}nyi networks
    seem to be the least vulnerable and the second most efficient. Ring
    substrates are both vulnerable and inefficient for these levels.
    ( X : small world networks, $\Delta$ :  ring substrates, O :
    scale free networks, $+$ : Erd\"{o}s-R\'{e}nyi networks )}
    \label{fig_VL}
\end{figure*}

\begin{figure*}[!tbp]
        \subfloat{\includegraphics[height=4.4cm]{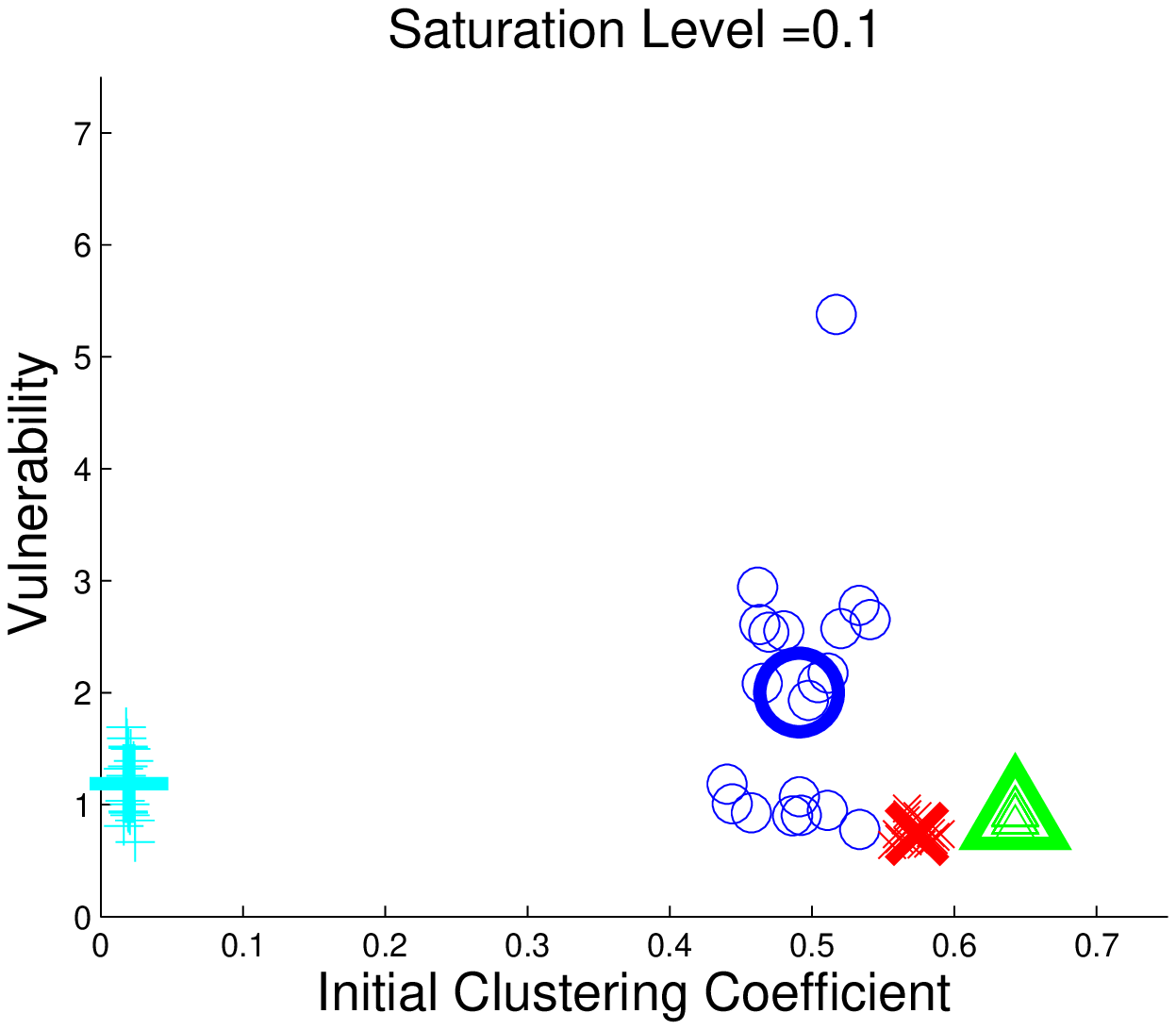}}
        \subfloat{\includegraphics[height=4.4cm]{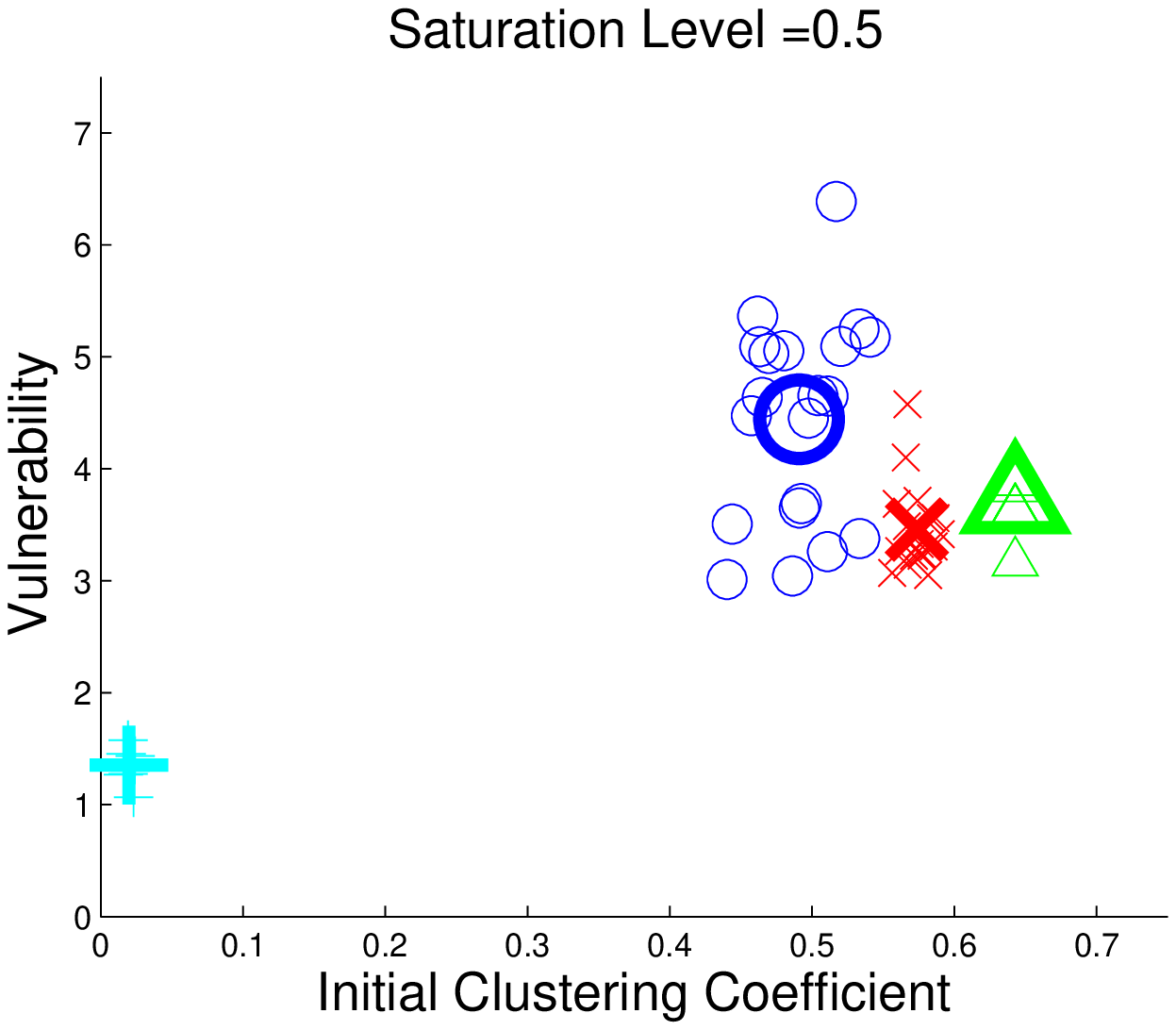}}
        \subfloat{\includegraphics[height=4.4cm]{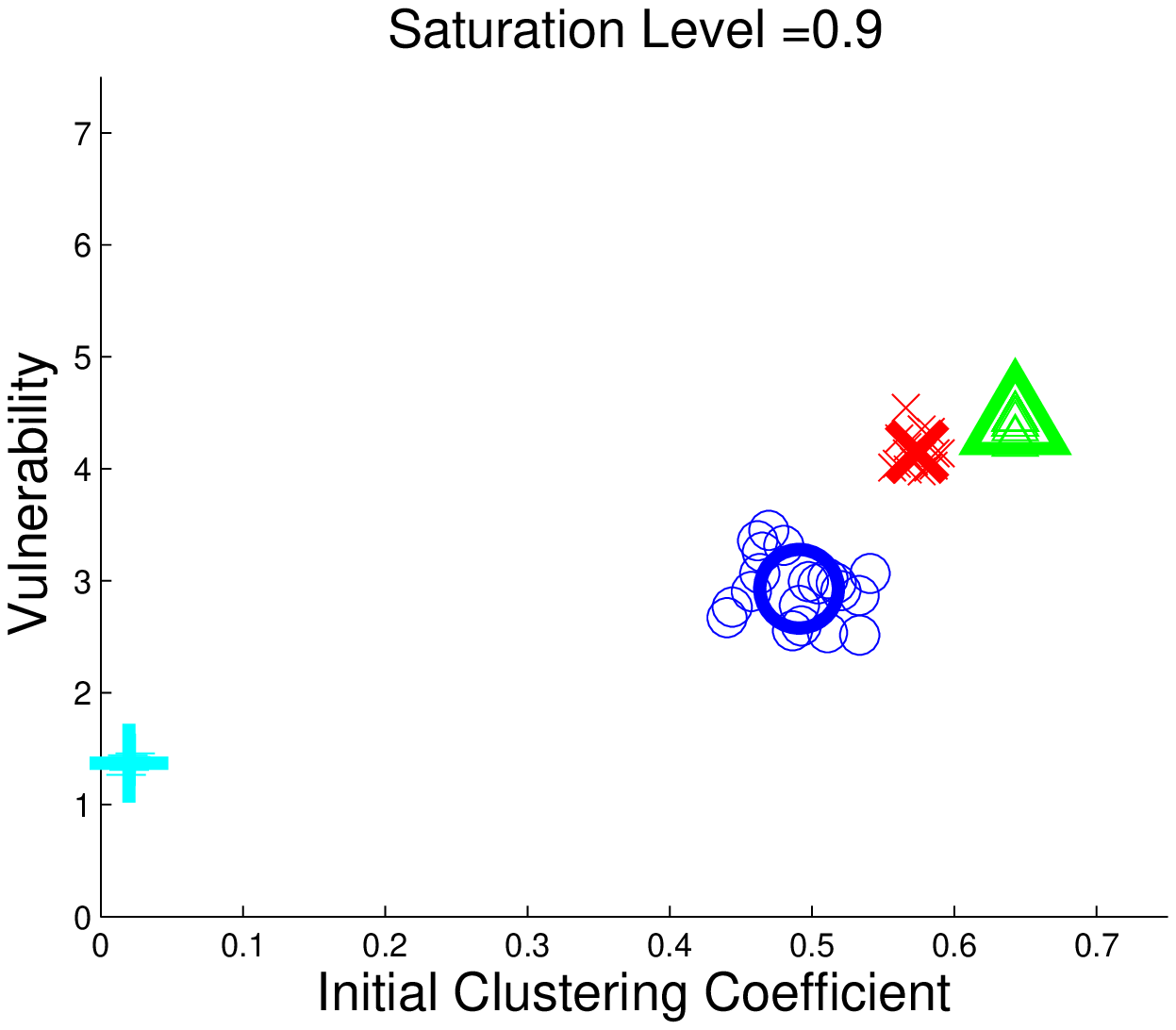}}
    \caption{Vulnerability versus initial clustering
    coefficient for the four network types and saturation levels of 0.1,
    0.5 and 0.9. Smaller markers depict results of a single network realization,
    and the bigger markers represent the mean points of
    such realizations. Erd\"{o}s-R\'{e}nyi networks have very small
    clustering coefficients but low vulnerability as well. When
    compared to small world networks and ring substrates, scale free
    networks are less vulnerable for smaller saturation levels, but
    become more vulnerable with increasing saturation level. ( X : small world
    networks, $\Delta$ :  ring substrates, O : scale free networks,
    $+$ : Erd\"{o}s-R\'{e}nyi networks )} \label{fig_VC}
\end{figure*}

Networks respond differently to varying levels of link failures,
therefore it may be more informative to calculate the vulnerability
of a network for a range of link failure ratios. The
differences resulting from the subjective evaluation of a saturation
level can also yield beneficial information. Here we use three
saturation levels of 10, 50 and 90 percent node pair disconnections. 
These saturation levels represent the possible states of a network
after failures occur. If the subject at hand is a highway network in
which bridges are prone to collapse in case of an earthquake, the
expected damage is low and short-term, which suggests choosing a
lower saturation level. On the other hand if a failure is accepted
to be congestion on a highway, then large, long-term, cascading
failures are expected. Thus choosing a high saturation level is 
more appropriate. Figure \ref{satur} shows three disconnection cases
that correspond to the saturation levels.

It is important to note the dependency of the vulnerability value on
the selected saturation level, for slight changes in saturation
levels result in different vulnerability values. Hence in order to
properly decide on a saturation level, one must carefully consider
the true function of the network and the failure scenarios for which
the network is being tested against.

When comparing networks in these terms and bearing in mind the
results obtained, it intuitively seems that improving
network efficiency without increasing the average degree can only be
achieved by removing short range links at the local level and adding
long range links connecting formerly distant communities. However
this would also cause increased vulnerability since redundancies
would be removed. To utilize this information for selecting a proper
network type, the initial clustering coefficient and efficiency of
the network must be taken into account. One might consider vulnerability
as a measure that can be used to relate the efficiency and the
clustering coefficient of a network, or simply a third dimension on which
different network types can be further compared.

It can be seen from Figure \ref{fig_VL} that scale free networks are
more efficient when compared to others, but always very vulnerable
for small saturation levels. For higher saturation levels, however,
scale free networks are not as vulnerable as rings or small worlds.
It should be noted that Erd\"{o}s-R\'{e}nyi networks seem to be
efficient above average and resistant to various saturation levels,
therefore should be the best choice if the prime concern is to have
a network that is both resistant to failures and efficient.

Figure \ref{fig_VC} shows, on the other hand, that if a clustered
network is preferred, Erd\"{o}s-R\'{e}nyi networks are the worst
choice despite their low vulnerability values. Performances of the
other network types are very similar in terms of clustering, but
they differ in terms of vulnerability for different saturation
levels. For the 0.1 and 0.5 levels, scale free networks are more
vulnerable then the others. For the 0.9 level, however, scale free
networks are less vulnerable and less clustered when compared to
small world networks and ring substrates.

\section{CONCLUSION}

The aim of this study was to discuss how some of the well-known
network types perform when they suffer link failures. To this end,
various methods of ranking were used to decide on the importance of
the links in the networks, and failures were carried out by removing
the links starting with the highest rated one. Based on the result of
the simulations, a very simple vulnerability index was defined and
it was observed that this index is capable of differentiating the
performance of the networks for different saturation levels.

The results obtained show that in terms of long term vulnerability,
ring structures are the most vulnerable amongst the network types
investigated herein, followed by small worlds. On the other hand,
for relatively higher average degrees, small worlds appear to be
closer to Erd\"{o}s-R\'{e}nyi and scale-free networks in terms of
vulnerability; they are redundant on the local level despite being
less efficient. In terms of short term vulnerability, however, the
difference between small worlds and more random networks disappear;
moreover, small worlds become less vulnerable at smaller saturation
levels. Thus if the subject network is prone to failures of bigger
magnitudes and if conditions permit, a power-law distribution would
be a wiser choice for an infrastructure network in terms of local
redundancies, vulnerability and efficiency. If, however, there is
only a negligible probability of mass failures, then the
vulnerability of small worlds and random networks are more or less
the same; in such cases the function of the network would determine
whether higher clustering is to be preferred over higher efficiency
or vice versa.

\bibliographystyle{unsrt}
\bibliography{LaTeX1}

\begin{thebibliography}{10}

\bibitem{vulnera1}
R\'eka Albert, Hawoong Jeong, and Albert~Laszlo Barabasi.
\newblock {Error and attack tolerance of complex networks}.
\newblock {\em Nature}, 406(6794):378--382, July 2000.

\bibitem{vulnera2}
P.~Holme, B.~J. Kim, C.~N. Yoon, and S.~K. Han.
\newblock {Attack vulnerability of complex networks}.
\newblock {\em Phys. Rev. E}, 65(5):056109+, May 2002.

\bibitem{vulnera3}
Vito Latora and Massimo Marchiori.
\newblock {Vulnerability and Protection of Critical Infrastructures}.
\newblock July 2004.

\bibitem{vulinf0}
Luca Dall'asta, Alain Barrat, Marc Barthelemy, and Alessandro Vespignani.
\newblock {Vulnerability of weighted networks}.
\newblock Mar 2006.

\bibitem{vulinf1}
B.~Berche, C.~von Ferber, T.~Holovatch, and Y.~Holovatch.
\newblock {Resilience of public transport networks against attacks}.
\newblock {\em The European Physical Journal B - Condensed Matter and Complex
  Systems}, 71(1):125--137, Sep 2009.

\bibitem{vulinf2}
D.~Chassin and C.~Posse.
\newblock {Evaluating North American electric grid reliability using the
  Barabási–Albert network model}.
\newblock {\em Physica A: Statistical Mechanics and its Applications},
  355(2-4):667--677, Sep 2005.

\bibitem{vulinf3}
Ricard~V. Sol\'e, Mart\'\i{} Rosas-Casals, Bernat Corominas-Murtra, and Sergi
  Valverde.
\newblock Robustness of the european power grids under intentional attack.
\newblock {\em Phys. Rev. E}, 77(2):026102, Feb 2008.

\bibitem{vulinf4}
R\'eka Albert, Istv\'an Albert, and Gary~L. Nakarado.
\newblock Structural vulnerability of the north american power grid.
\newblock {\em Phys. Rev. E}, 69(2):025103, Feb 2004.

\bibitem{vulinf5}
Rui Carvalho, Lubos Buzna, Flavio Bono, Eugenio Guti\'errez, Wolfram Just, and
  David Arrowsmith.
\newblock Robustness of trans-european gas networks.
\newblock {\em Phys. Rev. E}, 80(1):016106, Jul 2009.

\bibitem{cent0}
Karen Stephenson and Marvin Zelen.
\newblock Rethinking centrality: Methods and examples.
\newblock {\em Social Networks}, 11(1):1 -- 37, 1989.

\bibitem{cent2}
Stephen~P. Borgatti.
\newblock Centrality and network flow.
\newblock {\em Social Networks}, 27(1):55 -- 71, 2005.

\bibitem{cent3}
Stephen~P. Borgatti and Martin~G. Everett.
\newblock A graph-theoretic perspective on centrality.
\newblock {\em Social Networks}, 28(4):466 -- 484, 2006.

\bibitem{btwcent0}
Linton~C Freeman.
\newblock A set of measures of centrality based upon betweenness.
\newblock {\em Sociometry}, 40:35--41, 1977.

\bibitem{btwcent1}
M.~Barthélemy.
\newblock Betweenness centrality in large complex networks.
\newblock {\em The European Physical Journal B - Condensed Matter and Complex
  Systems}, 38:163--168, 2004.
\newblock 10.1140/epjb/e2004-00111-4.

\bibitem{rw}
Jae~Dong Noh and Heiko Rieger.
\newblock Random walks on complex networks.
\newblock {\em Phys. Rev. Lett.}, 92(11):118701, Mar 2004.

\bibitem{rwcent}
M.E.~J. Newman.
\newblock A measure of betweenness centrality based on random walks.
\newblock {\em Social Networks}, 27(1):39 -- 54, 2005.

\bibitem{btwrwcent}
M.~E.~J. Newman and M.~Girvan.
\newblock Finding and evaluating community structure in networks.
\newblock {\em Phys. Rev. E}, 69(2):026113, Feb 2004.

\bibitem{sf}
A.~L. Barabasi and R.~Albert.
\newblock {Emergence of Scaling in Random Networks}.
\newblock {\em Science}, 286(5439):509--512, October 1999.

\bibitem{graph4}
M.~L. Goldstein, S.~A. Morris, and G.~G. Yen.
\newblock {Problems with fitting to the power-law distribution}.
\newblock {\em The European Physical Journal B - Condensed Matter and Complex
  Systems}, 41(2):255--258, September 2004.

\bibitem{graph0}
Michael Molloy and Bruce Reed.
\newblock {A critical point for random graphs with a given degree sequence}.
\newblock {\em Random Structures \& Algorithms}, 6:161--180, 1995.

\bibitem{graph2}
R.~Milo, N.~Kashtan, S.~Itzkovitz, M.~E.~J. Newman, and U.~Alon.
\newblock {On the uniform generation of random graphs with prescribed degree
  sequences}.
\newblock May 2004.

\bibitem{sw0}
D.~J. Watts and S.~H. Strogatz.
\newblock {Collective dynamics of 'small-world' networks.}
\newblock {\em Nature}, 393(6684):440--442, June 1998.

\bibitem{sw1}
Jon Kleinberg.
\newblock The small-world phenomenon: an algorithm perspective.
\newblock In {\em Proceedings of the thirty-second annual ACM symposium on
  Theory of computing}, STOC '00, pages 163--170, New York, NY, USA, 2000. ACM.

\bibitem{sw2}
M.~E.~J. Newman.
\newblock {Models of the Small World: A Review}.
\newblock May 2000.

\bibitem{fragratio}
M.C. González, H.J. Herrmann, J.~Kertész, and T.~Vicsek.
\newblock Community structure and ethnic preferences in school friendship
  networks.
\newblock {\em Physica A: Statistical Mechanics and its Applications},
  379(1):307 -- 316, 2007.

\bibitem{discnp}
M.~Gonz\'{a}lez, H.~Herrmann, and A.~Araujo.
\newblock {Cluster size distribution of infection in a system of mobile
  agents}.
\newblock {\em Physica A: Statistical Mechanics and its Applications},
  356(1):100--106, October 2005.

\bibitem{cc}
Duncan~J. Watts.
\newblock {\em {Small Worlds: The Dynamics of Networks between Order and
  Randomness (Princeton Studies in Complexity)}}.
\newblock Princeton University Press, illustrated edition edition, November
  2003.

\bibitem{eff}
L.~Dueñas-Osorio, J.~I. Craig, and B.~J. Goodno.
\newblock {Seismic response of critical interdependent networks}.
\newblock {\em Earthquake Engineering \& Structural Dynamics}, 36(2):285--306,
  September 2007.

\bibitem{tez}
S.~Colak.
\newblock Vulnerability of networks against rank ordered independent link
  failures.
\newblock Master's thesis, Bo\~gazi\c{c}i University, June 2010.

\bibitem{sim0}
S.~N. Dorogovtsev and J.~F.~F. Mendes.
\newblock {Evolution of networks}.
\newblock {\em Advances in Physics}, 51(4):1079--1187, 2002.

\bibitem{sim1}
R\'eka Albert and Albert-L\'aszl\'o Barab\'asi.
\newblock Statistical mechanics of complex networks.
\newblock {\em Rev. Mod. Phys.}, 74(1):47--97, Jan 2002.

\bibitem{sim2}
Guido Caldarelli and Alessandro Vespignani.
\newblock {\em {Large Scale Structure and Dynamics of Complex Networks: From
  Information Technology to Finance and Natural Science (Complex Systems and
  Interdisciplinary Science)}}.
\newblock World Scientific Publishing Company, June 2007.

\bibitem{sim3}
M.~E.~J. Newman.
\newblock Assortative mixing in networks.
\newblock {\em Phys. Rev. Lett.}, 89(20):208701, Oct 2002.

\bibitem{sim4}
M.~E.~J. Newman.
\newblock Mixing patterns in networks.
\newblock {\em Phys. Rev. E}, 67(2):026126, Feb 2003.

\bibitem{sim5}
Romualdo Pastor-Satorras, Alexei V\'azquez, and Alessandro Vespignani.
\newblock Dynamical and correlation properties of the internet.
\newblock {\em Phys. Rev. Lett.}, 87(25):258701, Nov 2001.

\bibitem{rel0}
David~R. Karger.
\newblock A randomized fully polynomial time approximation scheme for the all
  terminal network reliability problem.
\newblock In {\em Proceedings of the twenty-seventh annual ACM symposium on
  Theory of computing}, STOC '95, pages 11--17, New York, NY, USA, 1995. ACM.

\bibitem{rel1}
Daryl~D. Harms, Miro Kraetzl, Charles~J. Colbourn, and John~S. Devitt.
\newblock {\em Network Reliability: Experiments with a Symbolic Algebra
  Environment}.
\newblock CRC Press, Inc., Boca Raton, FL, USA, 1995.

\bibitem{rel2}
Michael~O. Ball, Charles~J. Colbourn, and J.S. Provan.
\newblock Network reliability.
\newblock 1992.

\end{thebibliography}

\end{document}